\documentclass[reprint,amsmath,amssymb,aps,]{revtex4-2}

\usepackage{tikz}
\usepackage{quantikz}
\usepackage{multirow}
\usepackage{graphicx}
\usepackage{tabularx} 
\usepackage{dcolumn}
\usepackage{tcolorbox}
\usepackage{xcolor} 
\usepackage{caption}
\usepackage{subcaption}
\usepackage{booktabs}
\usepackage{rotating} 
\usepackage{placeins} 
\usepackage{bm}
\usepackage{adjustbox}
\usepackage{bbold}
\usepackage{mathtools}
\usepackage{algpseudocode}
\usepackage{algpseudocode}
\usepackage[T1]{fontenc}
\usetikzlibrary{positioning, quotes}
\usepackage{todonotes}
\usepackage{booktabs}
\usepackage{siunitx}
\usepackage[backref=page]{hyperref}

\definecolor{BrickRed}{RGB}{178, 34, 34}        
\definecolor{ForestGreen}{RGB}{34, 139, 34}     
\definecolor{RoyalBlue}{RGB}{65, 105, 225}  

\begin{document}
\preprint{APS/123-QED}

\title{Cutting Slack: Quantum Optimization with Slack-Free Methods for Combinatorial Benchmarks}

\author{Monit Sharma$^{1}$}
\author{Hoong Chuin Lau$^{1,2}$}
\email{Corresponding author email: hclau@smu.edu.sg}
\address{$^1$School of Computing and Information Systems,  Singapore Management University, Singapore}
\address{$^2$Institute of High Performance Computing, A*STAR, Singapore}

\begin{abstract}
Constraint handling remains a key bottleneck in quantum combinatorial optimization. While slack-variable-based encodings are straightforward, they significantly increase qubit counts and circuit depth, challenging the scalability of quantum solvers. In this work, we investigate a suite of \emph{Lagrangian-based optimization techniques} including dual ascent, bundle methods, cutting plane approaches, and augmented Lagrangian formulations for solving constrained combinatorial problems on quantum simulators and hardware. Our framework is applied to three representative NP-hard problems: the Travelling Salesman Problem (TSP), the Multi-Dimensional Knapsack Problem (MDKP), and the Maximum Independent Set (MIS). 

We demonstrate that MDKP and TSP, with their inequality-based or degree-constrained structures, allow for slack-free reformulations, leading to significant qubit savings without compromising performance. In contrast, MIS does not inherently benefit from slack elimination but still gains in feasibility and objective quality from principled Lagrangian updates. We benchmark these methods across classically hard instances, analyzing trade-offs in qubit usage, feasibility, and optimality gaps. Our results highlight the flexibility of Lagrangian formulations as a scalable alternative to naive QUBO penalization, even when qubit savings are not always achievable. This work provides practical insights for deploying constraint-aware quantum optimization pipelines, with applications in logistics, network design, and resource allocation.

\end{abstract}

\maketitle

\section{\label{sec:level1} Introduction}

Combinatorial optimization lies at the heart of many real-world decision-making tasks in logistics, finance, network design, and operations research~\cite{du1998handbook,juan2015review,bengio2021machine,yu2000robust,zenios1993financial,resende2008handbook}. Solving these problems efficiently is particularly challenging due to their NP-hardness and the combinatorial explosion of feasible solutions.

Quantum computing has recently emerged as a promising framework for tackling such problems~\cite{hadfield2019quantum,Egger_2021,herrman2022multi,Nakanishi_2019,Higgott_2019,Barkoutsos_2020,crosson2021prospects,del2013shortcuts}. Methods such as Quantum Annealing (QA)~\cite{morita2008mathematical} and Variational Quantum Algorithms (VQAs)~\cite{peruzzo2014variational} aim to exploit quantum parallelism and entanglement to accelerate the search for optimal solutions. These algorithms typically rely on reformulating the original problem into a Quadratic Unconstrained Binary Optimization (QUBO) model, which can be mapped onto quantum hardware.

However, a central challenge in this reformulation lies in enforcing constraints. Most real-world problems, like the Multi-Dimensional Knapsack Problem (MDKP) \cite{mdkp} and Travelling Salesman Problem \cite{robinson1949hamiltonian,applegate2006traveling} (well-known NP-hard problems \cite{karp1972reducibility}) contain hard inequality constraints that cannot be directly encoded into a QUBO form. The standard workaround is to introduce \emph{slack variables}, transforming inequalities into equalities that can be penalized in the objective function~\cite{Lucas_2014}. While this technique simplifies constraint modeling, it introduces a substantial increase in the number of binary variables, thereby inflating qubit requirements and also deepening the circuit.

This slack-based approach introduces three major drawbacks:
\begin{itemize}
    \item \textbf{Qubit Overhead:}  Introducing slack variables increases the number of binary variables, leading to a higher qubit requirement that can make the problem infeasible for current quantum hardware.
    \item \textbf{Circuit Complexity:} Additional variables lead to deeper, noisier quantum circuits, impairing performance on Noisy Intermediate-Scale Quantum (NISQ) devices~\cite{preskill2018quantum}.
    \item \textbf{Optimization Challenges:} The enlarged search space can hinder convergence in variational solvers due to increased non-convexity and redundancy~\cite{sharma2025comparative}.
\end{itemize}

In this work, we explore a family of \emph{slack-free optimization strategies} that avoid the need for auxiliary slack variables altogether. More precisely, we explore the Lagrangian relaxation technique where we dualize the complicating constraints via Lagrange multipliers, which yields a sequence of relaxed subproblems, while the multipliers themselves are updated iteratively via a subgradient method~\cite{fisher1981lagrangian}. To improve feasibility and convergence, we investigate a range of dual optimization techniques, including Dual Averaging, Stochastic Subgradient, the Bundle Method, Cutting Plane Techniques, and Augmented Lagrangian method.

We experiment this approach on three combinatorial optimization problems: 
\begin{itemize}
    \item \textbf{Travelling Salesman Problem (TSP)}: We adopt the Held-Karp Lagrangian relaxation \cite{held1970traveling} to dualize degree constraints and reduce the reliance on slack variables. This allows us to encode the problem with significantly fewer qubits, by transforming hard constraints into soft penalties within the objective.
    \item \textbf{Multi-Dimensional Knapsack Problem (MDKP)}, which admits slack-free reformulations that offer significant qubit savings.
    \item \textbf{Maximum Independent Set (MIS)}, which lacks inequality constraints but still benefits from principled dual optimization techniques for improving feasibility and solution quality.
\end{itemize}

Our experimental pipeline evaluates these methods on realistic benchmark instances across simulators and quantum hardware. We compare qubit usage, feasibility, and optimality gaps across methods and analyze the trade-offs in convergence and scalability.

To the best of our knowledge, this work is the first to implement and systematically benchmark a range of Lagrangian-based dual update methods within a quantum gate-based optimization framework. Our findings offer practical guidance for implementing constraint-aware \cite{sharma2025adaptive} quantum optimization strategies that scale toward real-world deployment.

The remainder of this paper is structured as follows:  
 \textbf{Section~\ref{sec:level1}} reviews related work on constraint handling in quantum optimization.  
 \textbf{Section~\ref{sec:level2}} details the methodology for formulating MDKP using Lagrangian relaxation and unbalanced penalization.  
 \textbf{Section~\ref{sec:level3}} presents implementation details, including quantum encoding and optimization techniques. 
 \textbf{Section~\ref{sec:level4}} outlines the experimental evaluation, benchmark datasets, and performance metrics.  
 \textbf{Section~\ref{sec:discussion}} discusses the results, highlighting key insights and practical implications.  
Finally, we conclude with future research directions aimed at improving quantum constraint handling.

\section{\label{sec:level2}Literature Review}

Solving constrained combinatorial optimization problems on quantum platforms requires specialized constraint-handling strategies to ensure that quantum solvers can process them effectively. A widely adopted approach embeds constraints directly into the objective function using  penalty terms, transforming the problem into a Quadratic Unconstrained Binary Optimization (QUBO) formulation. However, conventional penalty-based methods often require large penalty coefficients or auxiliary slack variables, both of which pose computational challenges. Large penalties can dominate the objective, leading to poor quantum optimization performance, while slack variables increase the number of required qubits, limiting scalability \cite{karimi2017subgradientapproachconstrainedbinary,Monta_ez_Barrera_2023,Monta_ez_Barrera_2024, lee2025implementingslackfreecustompenalty}.

Recent research has explored alternative strategies that circumvent these limitations. These methods broadly fall into three categories:
\begin{enumerate}
    \item  \textbf{Slack-Based Approaches:} Introducing auxiliary variables to transform inequalities into equalities.
    \item  \textbf{Penalty-Based Methods:} Employing quadratic penalties or asymmetric penalty functions to penalize constraint violations without additional slack variables.
    \item  \textbf{Lagrangian Relaxation:} Reformulating constraints via dual variables and solving the relaxed problem iteratively.
\end{enumerate}

This section systematically reviews these approaches, comparing their theoretical underpinnings, empirical performance, and computational trade-offs.

\subsection{Slack-Based Approaches to Constraint Handling}

A standard technique for encoding inequality constraints in QUBO is the introduction of  slack variables. Given an inequality constraint:
\begin{equation}
    h(x) \leq 0,
\end{equation}
an auxiliary slack variable \( s \geq 0 \) is introduced such that:
\begin{equation}
    h(x) + s = 0.
\end{equation}
To encode \( s \) in binary form, it is expanded as:
\begin{equation}
    s = \sum_{k=0}^{m} 2^k s_k, \quad s_k \in \{0,1\}.
\end{equation}
where \( m \) determines the required precision.

While this transformation ensures exact constraint handling, it significantly increases the number of variables in the optimization problem. Empirical studies show that the slack-variable encoding can become infeasible for large instances. 

\subsection{Penalty-Based Methods}

An alternative to explicit slack variables is to  incorporate constraints as penalty terms into the objective function. The most common formulation penalizes constraint violations quadratically:
\begin{equation}
    \mathcal{L}(x) = f(x) + P \cdot h(x)^2.
\end{equation}
where \( P \) is a penalty coefficient. Theoretically, if \( P \) is chosen sufficiently large, constraint violations become too costly, ensuring feasibility. However, in practice, selecting an appropriate \( P \) is nontrivial. If \( P \) is too small, violations persist; if too large, the optimization landscape becomes ill-conditioned, making it difficult for quantum solvers to converge \cite{karimi2017subgradientapproachconstrainedbinary}.

To mitigate this, \cite{Monta_ez_Barrera_2023,Monta_ez_Barrera_2024} introduced  unbalanced penalization, which replaces a symmetric quadratic penalty with an asymmetric function:
\begin{equation}
    \zeta(x) = -\lambda_1 h(x) + \lambda_2 [h(x)]^2.
\end{equation}
where \( \lambda_1 \) and \( \lambda_2 \) are tunable parameters. The  linear term encourages feasibility by reducing the objective for valid solutions, while the  quadratic term imposes a penalty on violations.

\begin{figure*}[t]
    \centering
    \begin{subfigure}[t]{0.48\textwidth}
        \centering
        \includegraphics[width=\linewidth]{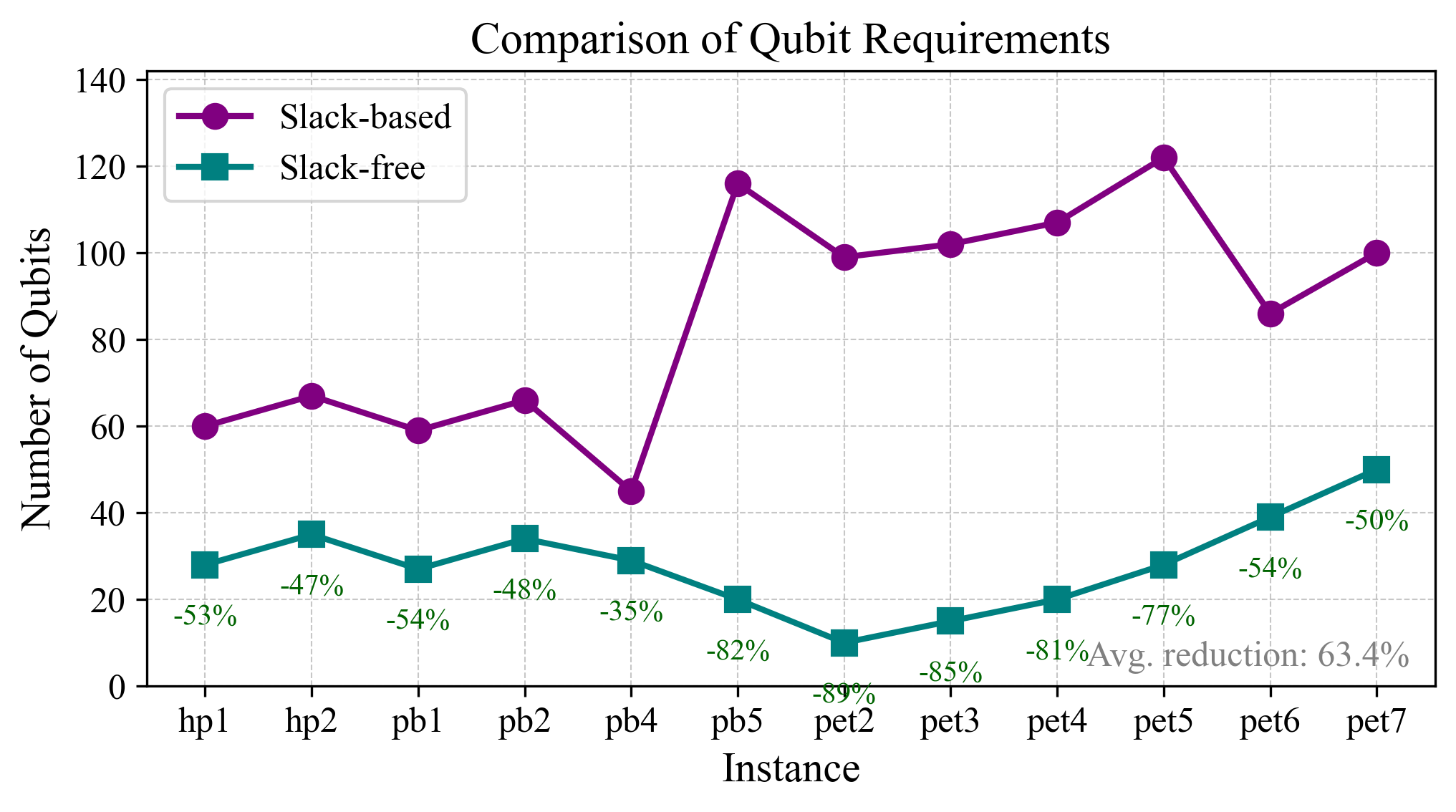}
        \label{fig:qubit_req}
    \end{subfigure}
    \hfill
    \begin{subfigure}[t]{0.48\textwidth}
        \centering
        \includegraphics[width=\linewidth]{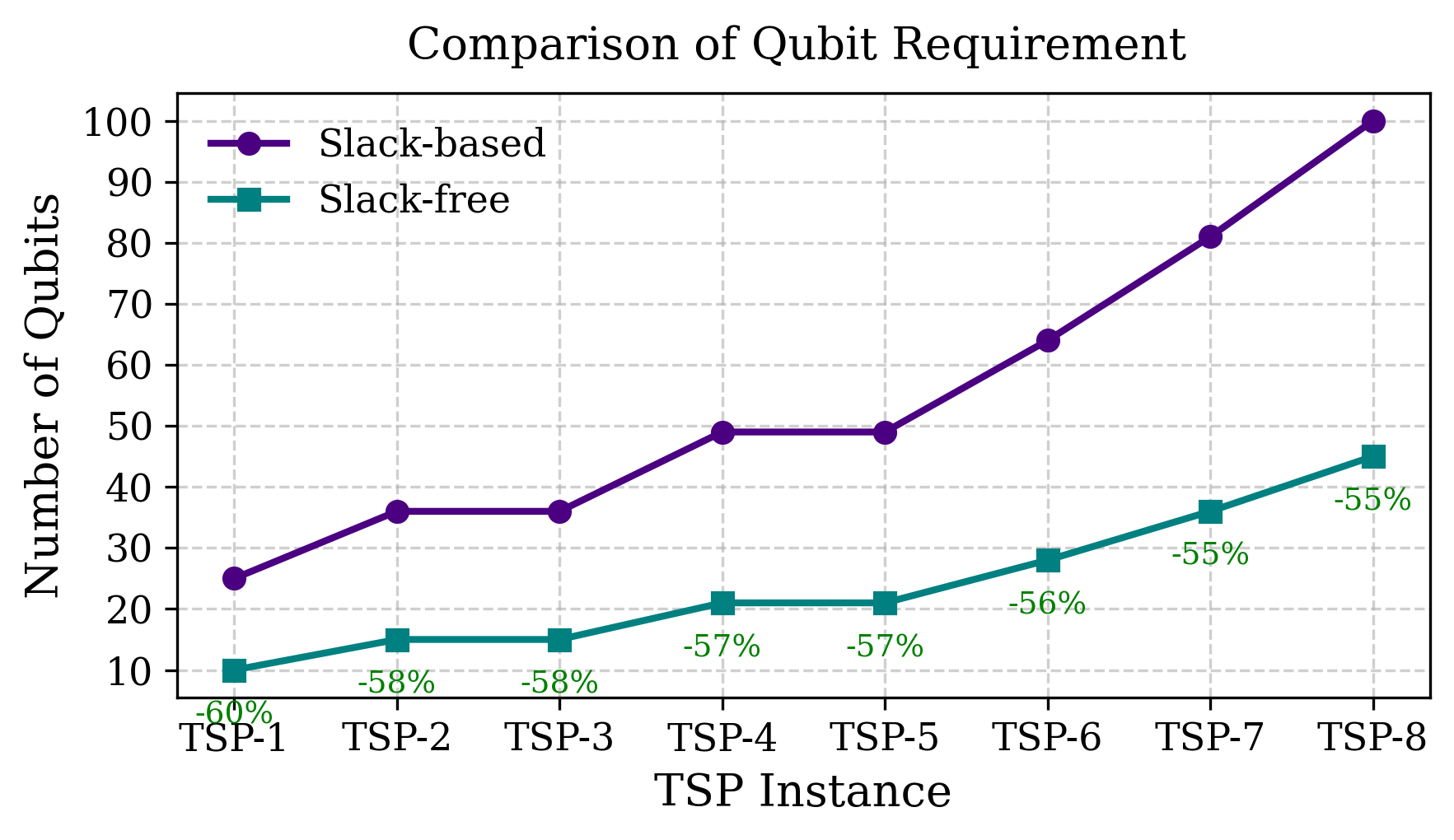}
        \label{fig:tsp_req}
    \end{subfigure}
    \caption{Comparison of qubit requirements across problem formulations. 
    \textbf{(a)} MDKP: Scaling of qubit requirements for slack-based (purple) vs slack-free (teal) formulations, demonstrating consistent savings from eliminating slack variables. 
    \textbf{(b)} TSP: Instance-wise comparison of slack-based (purple) and slack-free (teal) qubit counts, using the Held-Karp relaxation to avoid slack overhead.}
    \label{fig:qubit_comparison}
\end{figure*}

Empirical results indicate a clear scaling advantage for slack-free formulations over traditional penalty-based methods. As shown in Figure~\ref{fig:qubit_comparison}, the number of qubits needed by slack-based approaches increases steeply with problem size, whereas slack-free techniques maintain a more modest and manageable qubit requirement. This reduction in qubit overhead not only eases resource constraints on current quantum devices but also enables the efficient solution of larger combinatorial optimization problems. These findings underscore the potential of slack-free methods to extend the applicability of quantum optimization to more complex and larger-scale problems.

\subsection{Lagrangian Relaxation Approaches}

A more flexible strategy for handling constraints is  Lagrangian relaxation, 
which introduces  dual variables (Lagrange multipliers) to penalize violations of the complicating constraints and then iteratively updates those multipliers
to push the relaxed solutions toward feasibility.

More precisely, the constrained optimization problem:
\begin{equation}
    \min_x f(x) \quad \text{subject to} \quad h_m(x) \leq 0,
\end{equation}
is relaxed by defining the Lagrangian function:
\begin{equation}
    L(x, \lambda) = f(x) + \sum_{m} \lambda_m h_m(x),
\end{equation}
where \( \lambda_m \geq 0 \) are Lagrange multipliers. The  dual problem is then:
\begin{equation}
    \max_{\lambda \geq 0} \min_x L(x, \lambda).
\end{equation}

Several quantum optimization approaches have leveraged this approach:

\begin{itemize}
    \item A Lagrangian relaxation approach in quantum annealing was implemented by updating multipliers via a subgradient method:
    \begin{equation}
        \lambda_m^{(t+1)} = \max\{0, \lambda_m^{(t)} + \alpha^{(t)} h_m(x^{(t)})\},
    \end{equation}
as demonstrated in \cite{karimi2017subgradientapproachconstrainedbinary, ronagh2018solvingconstrainedquadraticbinary}.

    \item A Hubbard--Stratonovich transformation was adapted to convert quadratic penalties into auxiliary fields that enforce constraints stochastically, as presented in \cite{takabayashi2024subgradient}.

\item Lagrangian relaxation was integrated into a Variational Quantum Eigensolver (VQE) framework, where the variable \( x \) is optimized via quantum circuits while \( \lambda \) is adjusted classically, as demonstrated in \cite{le2024variationalquantumeigensolverconstraints}.

\end{itemize}

\subsection{Comparative Analysis}

Table~\ref{tab:constraint_methods} summarizes the key differences among these approaches.

\begin{table*}[h]
\centering
\caption{Comparison of Constraint Encoding Methods. The table compares different constraint-handling strategies in quantum optimization based on key computational characteristics.}
\label{tab:constraint_methods}
\renewcommand{\arraystretch}{1.2} 
\setlength{\tabcolsep}{10pt} 
\begin{tabular}{l c c c c}
\toprule
 \textbf{Method} &  \textbf{Qubit Overhead} &  \textbf{Feasibility Guarantee} &  \textbf{Computational Cost} &  \textbf{Scalability} \\
\midrule
 \textbf{Slack Variables} & \textcolor{BrickRed}{High} & \textcolor{ForestGreen}{Guaranteed} & \textcolor{RoyalBlue}{Moderate} & \textcolor{BrickRed}{Low}  \\
 \textbf{Penalty Methods} & \textcolor{ForestGreen}{None} & \textcolor{BrickRed}{Not Guaranteed} & \textcolor{RoyalBlue}{Low (single shot)}& \textcolor{RoyalBlue}{Moderate}  \\
 \textbf{Lagrangian Relaxation} & \textcolor{ForestGreen}{None} & \textcolor{RoyalBlue}{Strong (iterative)} & \textcolor{BrickRed}{High (multiple runs)} & \textcolor{ForestGreen}{High} \\
\bottomrule
\end{tabular}
\end{table*}

The reviewed methods provide distinct trade-offs in constraint handling for quantum optimization. Slack variable approaches guarantee feasibility—assuming a feasible solution exists and that the underlying QUBO solver finds an optimal solution—but they require a quadratic number of penalty terms, resulting in excessive qubit overhead. Penalty methods also involve a quadratic number of penalty terms but there is no qubit overhead; yet still does not guarantee strict feasibility. In contrast, Lagrangian relaxation employs only a linear number of penalty terms, making it more scalable and striking a balance between feasibility and resource efficiency. However, this benefit comes at the cost of iterative computation, potentially leading to high computational cost.

\subsection{Positioning Relative to Classical Methods}

While our review has focused on slack-based, penalty-based, and Lagrangian relaxation approaches within the quantum optimization context, it is important to situate these methods within the broader landscape of classical constraint-handling techniques. Classical optimization literature has long addressed the challenges of handling constraints in combinatorial problems such as the MDKP. For instance, techniques like cutting plane methods and Lagrangian relaxation (see, e.g., \cite{fisher1981lagrangian,pisinger1998knapsack, martello1988new,pisinger1999core}) have been extensively studied and successfully applied to handle complicating constraints while managing computational resources.

Classical solvers, however, typically rely on iterative linear programming or branch-and-bound frameworks that ensure feasibility only up to a limited problem size. In contrast, the slack-free formulations explored in this work aim to reduce qubit overhead on quantum platforms, thereby potentially scaling to larger instances of MDKP. Furthermore, while specialized quantum techniques—such as those employing tailored mixers in QAOA for enforcing hard constraints \cite{hadfield2019quantum}—offer alternative strategies, our approach is distinguished by its direct porting of classical Lagrangian relaxation and unbalanced penalization methods to a quantum framework.

A related line of work by Parjadis et al.~\cite{parjadis2023learning} explores the integration of classical Lagrangian relaxation into modern combinatorial optimization pipelines, specifically for the Travelling Salesman Problem (TSP). Their method leverages machine learning, particularly graph neural networks, to predict high-quality Lagrange multipliers for the Held-Karp relaxation, thereby accelerating the computation of dual bounds within classical constraint programming solvers. While their focus remains on enhancing classical pruning strategies via learned heuristics, our approach differs in its objective and execution: we directly port the Held-Karp Lagrangian relaxation into a slack-free quantum formulation, with the goal of minimizing qubit overhead rather than runtime within classical solvers. This contrast reinforces our broader motivation—to adapt and translate classical constraint-handling techniques into quantum-native frameworks that respect the limitations of near-term quantum hardware.

While classical solvers based on constraint programming and integer programming have demonstrated the ability to handle instances with thousands of items efficiently \cite{pisinger1998knapsack,pisinger1999core}, current quantum methods are typically limited to problems involving few qubits. This disparity in scalability underscores the urgent need for qubit-efficient formulations that can extend the reach of quantum optimization techniques to larger, more complex problem instances.

By explicitly contrasting our slack-free methods with these established classical techniques, we highlight the motivation behind our work: to bridge the gap between the scalability of classical constraint handling and the resource limitations inherent in current quantum hardware. This broader perspective not only reinforces the novelty of our approach but also preempts concerns regarding the omission of classical literature in our literature review.

\section{\label{sec:level3} Methodology}

This section outlines our methodological framework for solving constrained combinatorial optimization problems using quantum optimization. We begin by formulating each problem in its classical integer programming form and apply Lagrangian relaxation to handle hard constraints without introducing slack variables. This relaxation enables the construction of qubit-efficient QUBO representations suitable for quantum optimization. We then detail how these formulations are integrated with quantum solvers such as QAOA/VQE, and describe various dual update methods used to iteratively improve constraint satisfaction. The methodology is structured to reflect this pipeline: we first present the problem formulations, followed by the Lagrangian relaxation framework, the quantum integration strategy, and finally, the dual optimization techniques.

\subsection{Problem Formulations}

This section provides formal descriptions and Lagrangian relaxations for three combinatorial optimization problems that are central to our study: the Travelling Salesman Problem (TSP), the Multi-Dimensional Knapsack Problem (MDKP), and the Maximum Independent Set (MIS).

Each problem is first introduced in its standard integer programming form, followed by a derivation of its Lagrangian relaxation. The Lagrangian dual variables are interpreted to highlight their role in relaxing hard constraints, which enables more tractable subproblems and facilitates integration with quantum or hybrid optimization techniques. This formulation framework underpins the design and analysis of our quantum-classical algorithms in subsequent sections.

\subsubsection{Travelling Salesman Problem (TSP)}

The Travelling Salesman Problem (TSP) is a foundational problem in combinatorial optimization and operations research. It asks for the shortest possible route that visits each city exactly once and returns to the origin city. The TSP arises in logistics, circuit design, and computational biology, and is known to be NP-hard.

Let $G = (V, E)$ be a complete undirected graph with $|V| = n$ cities and edge weights $c_{ij} \geq 0$ denoting the travel cost between cities $i$ and $j$. Define binary variables $x_{ij} \in \{0, 1\}$, where $x_{ij} = 1$ if edge $(i,j)$ is in the tour.

The classical integer programming formulation is:

\begin{align}
\text{minimize} \quad & \sum_{i < j} c_{ij} x_{ij} \\
\text{subject to} \quad & \sum_{j \neq i} x_{ij} = 2, \quad \forall i \in V \label{eq:tsp-degree} \\
& x(S) \leq |S| - 1, \quad \forall S \subset V, \, 2 \leq |S| \leq n-1 \label{eq:tsp-subtour} \\
& x_{ij} \in \{0,1\}, \quad \forall (i,j) \in E
\end{align}

Constraint~\eqref{eq:tsp-degree} enforces that each node has degree 2, and constraint~\eqref{eq:tsp-subtour} eliminates subtours by ensuring connectivity. However, the subtour elimination constraints grow exponentially and are difficult to handle directly.

\paragraph{Held-Karp Lagrangian Relaxation.}

The Held-Karp relaxation \cite{held1970traveling} dualizes the degree constraints~\eqref{eq:tsp-degree} using Lagrange multipliers $\lambda_i \in \mathbb{R}$. The subproblem that emerges is a relaxation over 1-trees, which can be computed in polynomial time.

Define the Lagrangian as:

\begin{equation}
\mathcal{L}(x, \lambda) = \sum_{i < j} \left(c_{ij} - \lambda_i - \lambda_j\right) x_{ij} + 2 \sum_{i=1}^n \lambda_i
\end{equation}

This formulation modifies the edge costs as $\tilde{c}_{ij} = c_{ij} - \lambda_i - \lambda_j$. The Lagrangian subproblem then seeks a minimum-cost \emph{1-tree}, a spanning tree on $n - 1$ nodes plus two additional edges connecting the excluded node (typically node 1) back into the graph.

Let $T(\lambda)$ denote the optimal 1-tree under modified costs. The Lagrangian dual is then:

\begin{equation}
\max_{\lambda \in \mathbb{R}^n} \left[ \min_{x \in \mathcal{X}_{\text{1-tree}}} \mathcal{L}(x, \lambda) \right]
\end{equation}

where $\mathcal{X}_{\text{1-tree}}$ denotes the set of incidence vectors of all 1-trees on $n$ nodes. The Lagrangian value provides a valid lower bound on the optimal TSP tour cost.

\paragraph{Handling Subtours.} 
Subtours are not enforced explicitly but are penalized indirectly through the learned dual information. Any residual subtours in the quantum solution are detected via classical post-processing and addressed with either repair heuristics or additional cut-based iterations if needed.
\paragraph{Interpretation.} The dual variables $\lambda_i$ penalize nodes for having degrees different from two, thereby guiding the relaxed subproblem toward feasible tours. This relaxation is particularly powerful because the 1-tree subproblem is solvable in $O(n^2 \log n)$ time, offering tight bounds in branch-and-bound algorithms \cite{held1970traveling} and useful insights for hybrid classical-quantum optimization strategies.

\subsubsection{Multi-Dimensional Knapsack Problem (MDKP)}

The Multi-Dimensional Knapsack Problem (MDKP) is a generalization of the classical knapsack problem where multiple resource constraints must be satisfied simultaneously. It arises in diverse domains such as resource allocation, portfolio optimization, and logistics.

Let there be $n$ items and $m$ resource constraints. Each item $i$ has a profit $p_i$ and consumes $w_{ji}$ units of resource $j$. The capacity of resource $j$ is denoted by $c_j$. The binary decision variable $x_i \in \{0, 1\}$ indicates whether item $i$ is selected.

\begin{align}
\text{maximize} \quad & \sum_{i=1}^n p_i x_i \\
\text{subject to} \quad & \sum_{i=1}^n w_{ji} x_i \leq c_j, \quad \forall j = 1, \dots, m \\
& x_i \in \{0, 1\}, \quad \forall i = 1, \dots, n
\end{align}

To derive the Lagrangian relaxation, the resource constraints are dualized using non-negative multipliers $\lambda_j \geq 0$. The Lagrangian becomes:

\begin{equation}
\mathcal{L}(x, \lambda) = \sum_{i=1}^n p_i x_i - \sum_{j=1}^m \lambda_j \left( \sum_{i=1}^n w_{ji} x_i - c_j \right)
\end{equation}

Rewriting:

\begin{equation}
\mathcal{L}(x, \lambda) = \sum_{i=1}^n \left( p_i - \sum_{j=1}^m \lambda_j w_{ji} \right) x_i + \sum_{j=1}^m \lambda_j c_j
\end{equation}

The dual variables $\lambda_j$ adjust the effective profit of each item based on its resource usage, guiding the optimization towards feasibility. The Lagrangian relaxation leads to a problem that is separable across items and easier to optimize.

\subsubsection{Maximum Independent Set (MIS)}

The Maximum Independent Set (MIS) problem seeks the largest subset of vertices in a graph such that no two are adjacent. This problem is central to combinatorics and finds applications in scheduling, wireless communication, and register allocation.

Let $G = (V, E)$ be an undirected graph with $|V| = n$ nodes and $|E| = m$ edges. Define $x_i \in \{0, 1\}$ to indicate whether vertex $i$ is in the independent set.

\begin{align}
\text{maximize} \quad & \sum_{i=1}^n x_i \\
\text{subject to} \quad & x_i + x_j \leq 1, \quad \forall (i,j) \in E \\
& x_i \in \{0,1\}, \quad \forall i = 1, \dots, n
\end{align}

We relax the edge constraints using dual variables $\lambda_{ij} \geq 0$:

\begin{equation}
\mathcal{L}(x, \lambda) = \sum_{i=1}^n x_i - \sum_{(i,j) \in E} \lambda_{ij} (x_i + x_j - 1)
\end{equation}

This simplifies to:

\begin{equation}
\mathcal{L}(x, \lambda) = \sum_{i=1}^n \left( 1 - \sum_{j: (i,j) \in E} \lambda_{ij} \right) x_i + \sum_{(i,j) \in E} \lambda_{ij}
\end{equation}

The Lagrangian reduces the effective weight of selecting each node based on adjacency penalties, encouraging sparse and conflict-free subsets.

\subsection{Lagrangian Dual Formulation and Optimization}

Consider a generic binary optimization problem of the form:

\begin{align}
    \text{maximize} \quad & f(x) \\
    \text{subject to} \quad & Ax \leq b, \\
    & x \in \{0,1\}^n,
\end{align}

where \( A \in \mathbb{R}^{m \times n} \), \( b \in \mathbb{R}^m \), and \( f(x) \) is a linear or quadratic objective function.

To reduce constraint complexity, we relax the constraints \( Ax \leq b \) using Lagrange multipliers \( \lambda \geq 0 \), yielding the Lagrangian dual function:

\begin{equation}
    g(\lambda) = \lambda^\top b + \max_{x \in \{0,1\}^n} \left[ f(x) - \lambda^\top Ax \right].
\end{equation}

The corresponding dual problem becomes:

\begin{equation}
    \min_{\lambda \geq 0} g(\lambda).
\end{equation}

Since \( g(\lambda) \) is typically concave but may be non-differentiable, it is minimized using subgradient methods. At each iteration \( k \), the Lagrange multipliers are updated as:

\begin{equation}
    \lambda_j^{(k+1)} = \max\left(0, \lambda_j^{(k)} - t_k \cdot s_j^{(k)}\right),
\end{equation}

where the subgradient is computed as:

\begin{equation}
    s_j^{(k)} = b_j - \sum_{i=1}^n A_{ji} x_i^*(\lambda).
\end{equation}

Here, \( x^*(\lambda) \) denotes the solution to the relaxed subproblem at the current multipliers. This iterative refinement of \( \lambda \) progressively improves constraint satisfaction in the relaxed problem and provides useful dual information for guiding hybrid or penalized quantum optimization approaches.

\subsubsection{Integration of Lagrangian Relaxation with Quantum Solvers}

In our framework, we employ  Lagrangian relaxation to embed the constraints directly into the objective function through  Lagrange multipliers, \(\lambda\). Unlike traditional methods that solve the entire constrained problem on a quantum device, our approach  decouples the dual and primal updates. Specifically, we first optimize the dual problem using  classical subgradient-based methods to obtain a near-optimal set of multipliers. Once these multipliers are determined, they are reintegrated into the Lagrangian-relaxed objective, and the complete problem is reformulated as a  QUBO (or Ising Hamiltonian). This final formulation is then solved using a  quantum solver (e.g., via a Variational Quantum Eigensolver (VQE)). 

This  hybrid classical-quantum integration leverages the efficiency of classical optimization for the dual updates while harnessing quantum resources to tackle the final combinatorial subproblem.

The overall procedure is summarized in Algorithm~\ref{alg:lagrangian_integration} (see Appendix).

\subsubsection{Lagrangian Dual Optimization: Update Methods}

To efficiently solve the dual problem in Lagrangian relaxation, various update methods are employed to optimize the Lagrange multipliers \( \lambda \). These methods are crucial for ensuring convergence while balancing computational efficiency. The key update strategies include:

\begin{itemize}
    \item  \textbf{Dual Averaging Method}
    \item  \textbf{Stochastic Subgradient Optimization}
    \item  \textbf{Bundle Method}
    \item  \textbf{Cutting Plane Method}
    \item  \textbf{Augmented Lagrangian Method}
\end{itemize}

Each method is discussed in detail below.

\subsubsection*{Dual Averaging Method}

The  Dual Averaging Method \cite{dual1,dual2,dual3} updates the Lagrange multipliers using an average of all subgradients accumulated over iterations, stabilizing updates in noisy settings.

 Update Rule:
\begin{equation}
    \lambda^{k+1} = \max\{0,\, \lambda^k - \alpha_k \bar{s}^k\}
\end{equation}
where the  averaged subgradient is defined as:
\begin{equation}
    \bar{s}^k = \frac{1}{k} \sum_{t=1}^{k} s^t.
\end{equation}

Alternatively, maintaining a cumulative sum of subgradients:
\begin{equation}
    G^k = \sum_{t=1}^{k} s^t,
\end{equation}
yields the update rule:
\begin{equation}
    \lambda^{k+1} = \max\{0,\, -\alpha_k G^k\} \quad \text{(assuming } \lambda^0=0\text{)}.
\end{equation}

\subsubsection*{Stochastic Subgradient Method}

For large-scale problems with numerous constraints, computing the full subgradient at each iteration is computationally expensive. The  Stochastic Subgradient Method \cite{robbins1951stochastic,stochastic2} mitigates this by updating the multipliers using a random mini-batch of constraints.

 Stochastic Subgradient:
\begin{equation}
    \tilde{s}^k = \frac{1}{|\mathcal{B}_k|}\sum_{j\in\mathcal{B}_k} s_j^k
\end{equation}
where \( \mathcal{B}_k \) represents a randomly sampled subset of constraints.

 Update Rule:
\begin{equation}
    \lambda_j^{k+1} =
    \begin{cases}
    \max\{0,\, \lambda_j^k - \alpha_k \tilde{s}_j^k\}, & \text{if } j\in\mathcal{B}_k, \\
    \lambda_j^k, & \text{otherwise}.
    \end{cases}
\end{equation}

\subsubsection*{Bundle Method}

The  Bundle Method \cite{bundle1} constructs a local approximation of the dual function by maintaining a history of past subgradients and function values, refining the search direction and improving convergence stability.

 Algorithm Steps:
\begin{enumerate}
    \item Solve the  Lagrangian subproblem to compute the function value \( g_k \) and subgradient \( s_k \) at \( \lambda_k \).
    \item Store the triplet \( (\lambda_k, g_k, s_k) \) in a  bundle.
    \item Solve a  Quadratic Program (QP):
    \begin{equation}
        \min_{\lambda} \left\{ \hat{g}(\lambda) + \frac{1}{2} \|\lambda - \lambda_k\|^2 \right\}
    \end{equation}
    where \( \hat{g}(\lambda) \) is a piecewise linear model of the dual function.
    \item Accept the step if the function improves sufficiently; otherwise, take a  null step.
\end{enumerate}

\subsubsection*{Cutting Plane Method}

The  Cutting Plane Method \cite{cp1} iteratively refines a  piecewise-linear approximation of the dual function by adding linear constraints (cuts) based on subgradients, systematically narrowing down the feasible region.

 Algorithm Steps:
\begin{enumerate}
    \item Solve the  Lagrangian subproblem to evaluate the function value \( g_k \) and subgradient \( s_k \).
    \item Introduce a new linear constraint (cut):
    \begin{equation}
        g(\lambda) \geq g_k + s_k^\top (\lambda - \lambda_k).
    \end{equation}
    \item Solve a  Linear Program (LP) to determine \( \lambda_{k+1} \).
    \item Repeat until convergence criteria are met.
\end{enumerate}

\subsubsection*{Augmented Lagrangian Method}

The  Augmented Lagrangian Method \cite{al1,al2} enhances the standard Lagrangian by adding a quadratic penalty term to improve constraint handling and convergence.

 Augmented Lagrangian Function:
\begin{align}
    \mathcal{L}_A(x, \lambda, \mu) &= f(x) + \sum_{j=1}^{m} \lambda_j 
    \left( b_j - \sum_{i=1}^{n} w_{ji} x_i \right) \notag \\
    &\quad + \frac{\mu}{2} \sum_{j=1}^{m} 
    \left( b_j - \sum_{i=1}^{n} w_{ji} x_i \right)^2.
\end{align}

 Update Rules:
\begin{enumerate}
    \item  Primal Update: Minimize the augmented Lagrangian with respect to the primal variables \( x \):
    \begin{equation}
        x^{k+1} = \arg\min_{x} \mathcal{L}_A(x, \lambda^k, \mu^k).
    \end{equation}
    \item  Dual Update: Adjust the Lagrange multipliers based on constraint violations:
    \begin{equation}
        \lambda_j^{k+1} = \lambda_j^k + \mu^k \left( b_j - \sum_{i=1}^{n} w_{ji} x_i^{k+1} \right).
    \end{equation}
    \item  Penalty Parameter Update: Optionally, update \( \mu \) to balance contributions from multipliers and penalties.
\end{enumerate}

Table~\ref{tab:update_methods} provides a comprehensive comparison of different update strategies employed in Lagrangian dual optimization. Each method is evaluated based on its update rule, advantages, and limitations, highlighting their suitability for various problem settings. The table underscores the trade-offs between computational efficiency, convergence stability, and complexity in solving large-scale constrained optimization problems.

\renewcommand{\arraystretch}{1.3} 
\begin{table*}[h]
\caption{Comparison of Lagrangian Dual Optimization Update Methods}
\label{tab:update_methods}
\centering
\resizebox{\textwidth}{!}{ 
\begin{tabular}{|m{4cm}|m{8cm}|m{4.5cm}|m{4.5cm}|}
\hline
 \textbf{Method} &  \textbf{Update Rule} &  \textbf{Advantages} &  \textbf{Disadvantages} \\
\hline

\multirow{3}{*}{ \textbf{Dual Averaging}} &
\parbox{7cm}{\raggedright
\begin{align*}
\lambda^{k+1} &= \max\{0, \lambda^k - \alpha_k \bar{s}^k\}, \quad 
\bar{s}^k = \frac{1}{k} \sum_{t=1}^{k} s^t,  \\
G^k &= \sum_{t=1}^{k} s^t, \quad 
\lambda^{k+1} = \max\{0, -\alpha_k G^k\} \quad (\text{if } \lambda^0=0).
\end{align*}
} &
\begin{minipage}{4cm}
\begin{itemize}
    \item Stabilizes updates, especially in noisy settings.
    \item Simple and easy to implement.
\end{itemize}
\end{minipage} &
\begin{minipage}{4cm}
\begin{itemize}
    \item Requires careful tuning of step size \( \alpha_k \).
    \item Convergence can be slow for large-scale problems.
\end{itemize}
\end{minipage} \\
\hline

\multirow{2}{*}{ \textbf{Stochastic Subgradient}} &
\parbox{7cm}{\raggedright
\begin{align*}
\lambda_j^{k+1} &=
\begin{cases}
\max\{0, \lambda_j^k - \alpha_k \tilde{s}_j^k\}, & j\in\mathcal{B}_k, \\
\lambda_j^k, & \text{otherwise}.
\end{cases}
\end{align*}
\begin{equation*}
\tilde{s}^k = \frac{1}{|\mathcal{B}_k|}\sum_{j\in\mathcal{B}_k} s_j^k
\end{equation*}
} &
\begin{minipage}{4cm}
\begin{itemize}
    \item Reduces computational cost per iteration.
    \item Suitable for large-scale problems with many constraints.
\end{itemize}
\end{minipage} &
\begin{minipage}{4cm}
\begin{itemize}
    \item Introduces variance due to stochasticity.
    \item Requires selection of mini-batch size and step size.
\end{itemize}
\end{minipage} \\
\hline

\multirow{2}{*}{ \textbf{Bundle Method}} &
\parbox{7cm}{\raggedright
Maintains a bundle of past subgradients \( s_i \) and function values \( g_i \). Solves a quadratic program (QP):
\begin{align*}
\lambda_{k+1} = \arg\min_{\lambda} \Bigg\{ 
    \max_{i} \Big[ g_i + s_i^\top (\lambda - \lambda_i) \Big]  \\
    + \frac{\beta}{2} \|\lambda - \lambda_k\|^2  
\Bigg\}.
\end{align*}
where \( \beta \) stabilizes updates.
} &
\begin{minipage}{4cm}
\begin{itemize}
    \item Uses historical information for more accurate updates.
    \item Effective in handling non-smooth functions.
\end{itemize}
\end{minipage} &
\begin{minipage}{4cm}
\begin{itemize}
    \item Solving QP at each iteration can be computationally expensive.
    \item Requires storage of past subgradients.
\end{itemize}
\end{minipage} \\
\hline

\multirow{2}{*}{ \textbf{Cutting Plane}} &
\parbox{7cm}{\raggedright
Approximates the function using accumulated linear constraints (cuts):
\begin{equation*}
g(\lambda) \geq g_k + s_k^\top (\lambda - \lambda_k)
\end{equation*}
Solves an LP:
\begin{equation*}
\lambda_{k+1} = \arg\min_{\lambda} \left\{ \lambda \mid \lambda \text{ satisfies all cuts} \right\}
\end{equation*}
} &
\begin{minipage}{4cm}
\begin{itemize}
    \item Systematically refines the approximation of the dual function.
    \item Converges to the optimal solution by iteratively adding cuts.
\end{itemize}
\end{minipage} &
\begin{minipage}{4cm}
\begin{itemize}
    \item LPs can become large as more cuts are added.
    \item Computationally intensive for problems with many constraints.
\end{itemize}
\end{minipage} \\
\hline

\multirow{2}{*}{ \textbf{Augmented Lagrangian}} &
\parbox{7cm}{\raggedright
Adds a quadratic penalty to the Lagrangian:
\begin{align*}
\mathcal{L}_A(x, \lambda, \mu) &= f(x) + \sum_{j=1}^{m} \lambda_j \left( b_j - \sum_{i=1}^{n} w_{ji} x_i \right) \\
&\quad + \frac{\mu}{2} \sum_{j=1}^{m} \left( b_j - \sum_{i=1}^{n} w_{ji} x_i \right)^2.
\end{align*}

with iterative updates:
\begin{equation*}
\lambda_j^{k+1} = \lambda_j^k + \mu \left( b_j - \sum_{i=1}^{n} w_{ji} x_i^k \right)
\end{equation*}
} &
\begin{minipage}{4cm}
\begin{itemize}
    \item Enhances convergence by penalizing constraint violations.
    \item Balances between primal and dual updates.
\end{itemize}
\end{minipage} &
\begin{minipage}{4cm}
\begin{itemize}
    \item Requires careful tuning of penalty parameter \( \mu \).
    \item May involve solving complex subproblems in each iteration.
\end{itemize}
\end{minipage} \\
\hline

\end{tabular}}
\end{table*}

In our implementation, the subproblem is solved  classically by taking advantage of the  closed-form solution. The  quantum solver is applied only once, after the best Lagrange multipliers have been determined, to solve the resulting  QUBO formulation. This  hybrid classical-quantum workflow substantially reduces quantum overhead by decoupling the iterative dual updates from the final combinatorial optimization. As a result, the approach is rendered more practical for deployment on near-term quantum hardware.

\section{\label{sec:level4} Experimental Evaluation}

To evaluate the feasibility of our formulations, we utilized the  Qiskit \cite{qiskit2024} AerSimulator as our primary quantum computing framework for preliminary simulations. Classical computations were executed on a high-performance server powered by an Intel\textregistered~Xeon\textregistered~Gold 6154 CPU @ 3.00\,GHz, featuring 144 logical processors distributed across four sockets (18 cores per socket with two threads per core). For experimental validation, our formulations were deployed on Rigetti’s Ankaa‑3, an 84‑qubit universal gate‑model superconducting QPU accessed via AWS Braket. Circuit parameters were pre‑optimized using classical simulations and then directly implemented on the QPU without further fine‑tuning, thereby minimizing execution time and reducing the impact of quantum decoherence on computational accuracy.

Our goal is not to claim quantum advantage, but rather to assess how well slack-free Lagrangian formulations, when combined with quantum solvers, approximate the optimal solutions obtained by classical methods. Classical solvers typically resolve these instances in seconds, whereas our VQE-based quantum runs required several minutes, even hours on simulators, highlighting the trade-off between hardware constraints and formulation compactness.

\subsection{Benchmark Instances}
\label{sec:instances}

Our experimental evaluation is based on benchmark instances from three canonical combinatorial problems: TSP, MDKP, and MIS. These instances were selected or generated to span a range of sizes suitable for quantum and hybrid quantum-classical optimization.

For the TSP, we generated synthetic metric instances containing 5-10 cities using a custom Python script. Each instance was constructed by randomly sampling 2D coordinates uniformly over the range \([0,1000]\), followed by computing symmetric integer-valued Euclidean distance matrices. The resulting data was stored in standard TSPLIB format.

For the MDKP, we utilized the SAC-94 dataset from~\cite{dataset}, which consists of instances derived from real-world industrial problems and has been widely used for benchmarking knapsack-based formulations.

To evaluate our approach on the MIS problem, we employed well-established benchmark graphs sourced from datasets based on error-correcting codes~\cite{mis_datasets}. These instances are known for their combinatorial complexity and are commonly used to assess the effectiveness of optimization algorithms for the Maximum Independent Set problem.

\subsection{Performance Metrics}

The performance of the proposed approach was evaluated using the following metrics:
\begin{itemize}
    \item  \textbf{Optimality Gap (\%)}: The percentage difference between the obtained solution and the known optimal solution.
\begin{align}
\text{Opt. Gap (\%)} = 
    & \frac{
    \text{Obj. Best} 
    - \text{Obj. Obtained}
    }{
    \text{Obj. Best}
    } \nonumber 
    & \times 100
\end{align}

\item \textbf{Relative Solution Quality (\%):} This metric evaluates the quality of the obtained solution relative to the best-known or optimal solution. It is typically expressed as a percentage, calculated as:
    
    \begin{equation*}
        \text{RSQ (\%)} = \left( \frac{\text{Obj. Value of Obtained Solution}}{\text{Obj. Value of Best-Known Solution}} \right) \times 100
        \label{eq:rsq}
    \end{equation*}
    
    Higher values indicate solutions closer to the optimal, demonstrating the effectiveness of the algorithm or approach used.

    \item  \textbf{Qubit Utilization}: In the context of quantum computing approaches, this metric indicates the number of qubits required to represent the problem instance. It reflects the quantum resource efficiency of the algorithm.

\end{itemize}

\subsection{\textbf{Results}}
\label{sec:results}

This section presents the empirical performance of various constraint-handling formulations across the three combinatorial problems considered—Travelling Salesman Problem (TSP), Multi-Dimensional Knapsack Problem (MDKP), and Maximum Independent Set (MIS). Our results are summarized in Tables~\ref{tab:tsp-simulator-table}–\ref{tab:mis-hardware-table}, which report the optimization gap or relative solution quality (RSQ) alongside qubit requirements for each method.

We provide a comprehensive evaluation across MDKP, TSP, and MIS using both simulated and hardware-executed quantum runs. For MDKP, Table~\ref{tab:mdkp-simulator-table} reports simulator results, while Table~\ref{tab:mdkp-hardware-table} presents outcomes obtained on Rigetti’s Ankaa‑3 QPU. Similarly, for TSP and MIS, we include both simulator results (Tables~\ref{tab:tsp-simulator-table},~\ref{tab:mis-simulator-table}) and hardware evaluations (Tables~\ref{tab:tsp-hardware-table},~\ref{tab:mis-hardware-table}). In the case of MIS, relative solution quality (RSQ) is used instead of an optimization gap, as exact optima are known. Across all problems, our slack-free methods demonstrate consistent and substantial qubit savings, making them well-suited for execution on near-term quantum hardware.

\subsubsection*{\textbf{Interpretation of Table Columns}}

\begin{itemize}
    \item \textbf{Instance:} The benchmark instance name corresponding to TSP, MDKP, or MIS.
    \item \textbf{Qubits (S/NS):} The number of qubits used for slack-based (S) and slack-free (NS) formulations. The S formulation introduces ancilla/slack variables to handle inequality constraints, while NS avoids these by encoding constraints via Lagrangian multipliers.
    \item \textbf{Slack QUBO:} Optimization gap obtained from solving the slack-based QUBO formulation directly.
    \item \textbf{Cutting Plane:} Performance of the cutting-plane-enhanced formulation, where constraint violations iteratively generate new inequalities.
    \item \textbf{Subgrad:} Results from the stochastic subgradient-based Lagrangian dual optimization.
    \item \textbf{Bundle:} A refinement of subgradient methods using memory of past gradients to improve stability.
    \item \textbf{Dual:} Dual Averaging approach using accumulated gradients to guide dual updates.
    \item \textbf{Aug. Lag.:} The Augmented Lagrangian method that introduces a quadratic penalty term for stronger constraint violation penalization.
    \item \textbf{--:} Indicates infeasibility due to solver failure or excessive runtime.
    \item \textbf{Q.L:} Indicates that the qubit limit of the quantum hardware was exceeded for that instance.
\end{itemize}

\begin{table*}[h]
\centering
\caption{\label{tab:tsp-simulator-table}Optimization gaps (\%) and qubit usage for TSP instances solved via quantum simulation using Slack QUBO, Cutting Plane, Subgradient, Bundle, and Dual method. Best results per instance are bolded.}
\begin{ruledtabular}
\begin{tabular*}{\textwidth}{@{\extracolsep{\fill}}lcccccc}
\textbf{Instance} & \textbf{Qubits (S/NS)} & \textbf{Slack QUBO} & \textbf{Cutting Plane} & \textbf{Subgrad} & \textbf{Bundle} & \textbf{Dual}  \\
\hline
TSP-1  & 25/10 & 15.41 & 13.54 & \textbf{9.27} & 16.33 & 18.21   \\
TSP-2  & 36/15 & 25.05 & 15.97 & \textbf{10.68} & 16.58 & 18.22   \\
TSP-3  & 36/15 & 18.45 & 21.38 & \textbf{1.76} & 12.53 & 20.52  \\
TSP-4  & 49/21 & 6.18 & 11.74 & 8.29 & 7.26 & \textbf{5.69}   \\
TSP-5  & 49/21 & 10.36 & 14.31 & \textbf{7.09} & 13.31 & 11.32   \\
TSP-6  & 64/28 & 38.48 & 13.84 & \textbf{2.10} & 11.33 & 9.07   \\
TSP-7  & 81/36 & 22.08 & 21.36 & \textbf{6.75} & 8.29 & 7.62   \\
TSP-8  & 100/45 & 62.98 & 51.08 & 39.89 & \textbf{29.32} & 60.52   \\
\end{tabular*}
\end{ruledtabular}
\end{table*}

\begin{table*}[h]
\centering
\caption{Optimization gaps (\%) and qubit usage for TSP instances on quantum hardware using Slack QUBO, Cutting Plane, Subgradient, Bundle, and Dual method. “–” = infeasible run, Q.L = exceeded qubit limit. Best result per instance is bolded.}
\label{tab:tsp-hardware-table}
\begin{ruledtabular}
\begin{tabular*}{\textwidth}{@{\extracolsep{\fill}}lcccccc}
\textbf{Instance} & \textbf{Qubits (S/NS)} & \textbf{Slack QUBO} & \textbf{Cutting Plane} & \textbf{Subgrad} & \textbf{Bundle} & \textbf{Dual}  \\
\hline
TSP-1  & 25 / 10  & 28.23 & \textbf{24.62} & 28.95 & 28.87 & 29.64 \\
TSP-2  & 36 / 15  & 39.76 & 28.05 & \textbf{23.41} & 29.22 & 31.01 \\
TSP-3  & 36 / 15  & 31.98 & 34.13 & \textbf{14.88}  & 28.05 & 32.78 \\
TSP-4  & 49 / 21  & 18.92  & 23.25 & 20.27         & 19.84  & \textbf{17.10} \\
TSP-5  & 49 / 21  & 23.77 & 26.88 & \textbf{19.61}  & 25.25 & 23.29 \\
TSP-6  & 64 / 28  & 52.35 & 26.41 & \textbf{14.02}  & 24.17 & 20.63 \\
TSP-7  & 81 / 36  & -- & 33.91 & \textbf{18.78}  & 20.04 & 19.71 \\
TSP-8  & 100 / 45 & Q.L   & 65.62 & 54.22         & \textbf{45.50} & 74.88 \\
\end{tabular*}
\end{ruledtabular}
\end{table*}

\begin{table*}[h]
\centering
\caption{\label{tab:mdkp-simulator-table}Optimization gaps (\%) and qubit usage for MDKP instances solved on simulator using various methods: Slack-based QUBO, Cutting Plane, Subgradient (stochastic), Bundle, Dual, and Augmented Lagrangian. Best performance per row is highlighted. ``--'' indicates infeasible runs.}
\begin{ruledtabular}
\begin{tabular*}{\textwidth}{@{\extracolsep{\fill}}lccccccc}
\textbf{Instance} & \textbf{Qubits (S/NS)} & \textbf{Slack QUBO} & \textbf{Cutting Plane} & \textbf{Subgrad} & \textbf{Bundle} & \textbf{Dual} & \textbf{Aug. Lag.} \\
\hline
hp1  & 60/28  & 39.76  & 17.02 & \textbf{9.71}   & 17.08  & 17.98 & 22.49 \\
hp2  & 67/35  & 12.34  & 29.88 & \textbf{10.16} & 30.91  & 15.37 & 15.75 \\
pb1  & 59/27  & 19.94  & 24.59 & \textbf{11.52} & 28.41  & 16.14 & 14.36 \\
pb2  & 66/34  & 19.49  & 30.72 & \textbf{15.97} & 19.17  & 20.62 & 28.18 \\
pb4  & 45/29  & --     & --    & 38.08         & 33.33  & 63.62 & \textbf{37.03} \\
pb5  & 116/20 & \textbf{4.25} & 27.52 & 7.05          & 38.01  & 14.11 & 43.15 \\
pet2 & 99/10  & 41.07  & \textbf{4.24}  & \textbf{4.24} & 18.12  & \textbf{4.24} & 13.41 \\
pet3 & 102/15 & 4.98   & 9.33  & \textbf{4.23} & \textbf{4.23} & 17.43 & 32.50 \\
pet4 & 107/20 & 66.58  & 19.60 & 30.14         & 22.05  & \textbf{18.46} & 44.93 \\
pet5 & 122/28 & 33.23  & 29.19 & 20.40         & 24.67  & \textbf{18.34} & 26.77 \\
pet6 & 86/39  & \textbf{12.50} & 21.82 & 25.75         & 19.68  & 22.44 & 50.42 \\
pet7 & 100/50 & 43.46  & 20.34 & \textbf{13.52} & 23.20  & 19.07 & 15.71 \\
\end{tabular*}
\end{ruledtabular}
\end{table*}

\begin{table*}[h]
\centering
\caption{\label{tab:mdkp-hardware-table}Optimization gaps (\%) and qubit usage for MDKP instances on quantum hardware using Slack QUBO, Cutting Plane, Subgradient (best of stochastic/explicit), Bundle, Dual, and Augmented Lagrangian methods. ``--'' = infeasible run, Q.L = exceeded qubit limit. Best result per instance is bolded.}
\begin{ruledtabular}
\begin{tabular*}{\textwidth}{@{\extracolsep{\fill}}lccccccc}
\textbf{Instance} & \textbf{Qubits (S/NS)} & \textbf{Slack QUBO} & \textbf{Cutting Plane} & \textbf{Subgrad} & \textbf{Bundle} & \textbf{Dual} & \textbf{Aug. Lag.} \\
\hline
hp1  & 60/28  & 42.10  & 17.32  & \textbf{20.88}  & \textbf{17.11} & 28.26 & 40.87 \\
hp2  & 67/35  & \textbf{21.78}  & 33.92  & 24.63  & 37.31 & 31.60 & 33.52 \\
pb1  & 59/27  & 22.52  & 35.79  & \textbf{15.72}  & 33.94 & 18.60 & 29.44 \\
pb2  & 66/34  & 22.70  & 28.84  & \textbf{10.70}  & 34.36 & 53.35 & 32.58 \\
pb4  & 45/29  & --     & --     & --     & \textbf{54.51} & --    & -- \\
pb5  & 116/20 & Q.L    & 64.23  & \textbf{43.75}  & 48.48 & 46.33 & -- \\
pet2 & 99/10  & Q.L    & 89.31  & \textbf{16.85}  & --    & 17.17 & 44.35 \\
pet3 & 102/15 & Q.L    & 68.86  & \textbf{20.42}  & 30.01 & 40.97 & 34.74 \\
pet4 & 107/20 & Q.L    & 43.70  & 33.66  & \textbf{26.06} & 40.93 & 52.04 \\
pet5 & 122/28 & Q.L    & 53.38  &  \textbf{26.77} & 30.40 & 33.75 & 31.85 \\
pet6 & 86/39  & Q.L    & 45.90  & \textbf{34.71}  & 46.53 & 40.26 & 51.77 \\
pet7 & 100/50 & Q.L    & 26.72  & \textbf{12.89}  & 50.51 & 24.55 & 31.05 \\
\end{tabular*}
\end{ruledtabular}
\end{table*}

\begin{table*}[h]
\centering
\caption{\label{tab:mis-simulator-table}Relative Solution Quality (\%) and qubit usage for MIS instances solved via quantum simulation using Slack QUBO, Cutting Plane, Subgradient, Bundle, Dual, and Augmented Lagrangian methods. The best result per instance is bolded.}
\begin{ruledtabular}
\begin{tabular*}{\textwidth}{@{\extracolsep{\fill}}lccccccc}
\textbf{Instance} & \textbf{Qubits (S/NS)} & \textbf{Slack QUBO} & \textbf{Cutting Plane} & \textbf{Subgrad} & \textbf{Bundle} & \textbf{Dual} & \textbf{Aug. Lag.} \\
\hline
1et64   & 64/64 & 77.8 & 77.8  & \textbf{88.89} & 77.8 & \textbf{88.89} & \textbf{88.89} \\
1tc8    & 8/8 & \textbf{100.0} & \textbf{100.0}  & \textbf{100.0} & \textbf{100.0} & \textbf{100.0} & \textbf{100.0} \\
1tc16   & 16/16 & 75.0 & 87.5 & 75.0 & 87.5 & 75.0 & \textbf{100.0} \\
1tc32   & 32/32 & 75.0 & 66.67 & \textbf{83.33} & 75.0 & 58.3 & \textbf{83.33} \\
1tc64   & 64/64 & 40.0 & 75.00 & 77.78& 40.0 & 50.0 & \textbf{90.0} \\
\end{tabular*}
\end{ruledtabular}
\end{table*}

\begin{table*}[h]
\centering
\caption{\label{tab:mis-hardware-table}Relative Solution Quality (\%) and qubit usage for MIS instances solved via quantum hardware using Slack QUBO, Cutting Plane, Subgradient, Bundle, Dual, and Augmented Lagrangian methods. The best result per instance is bolded. ``--'' = infeasible run, Q.L = exceeded qubit limit. }
\begin{ruledtabular}
\begin{tabular*}{\textwidth}{@{\extracolsep{\fill}}lccccccc}
\textbf{Instance} & \textbf{Qubits (S/NS)} & \textbf{Slack QUBO} & \textbf{Cutting Plane} & \textbf{Subgrad} & \textbf{Bundle} & \textbf{Dual} & \textbf{Aug. Lag.} \\
\hline
1et64   & 64/64 & -- & --  & 66.67 & -- & \textbf{77.78} & \textbf{77.78} \\
1tc8    & 8/8 & \textbf{100.0} & \textbf{100.0}  & \textbf{100.0} & \textbf{100.0} & \textbf{100.0} & \textbf{100.0} \\
1tc16   & 16/16 & 50.0 & 75.0 & 50.0 & -- & -- & \textbf{87.5} \\
1tc32   & 32/32 & -- & -- & \textbf{33.33} & \textbf{33.33} & -- & \textbf{33.33} \\
1tc64   & 64/64 & -- & -- & -- & -- & -- & -- \\
\end{tabular*}
\end{ruledtabular}
\end{table*}

\begin{figure*}[htbp!]
    \centering
    \begin{subfigure}[b]{0.48\linewidth}
        \includegraphics[width=\linewidth]{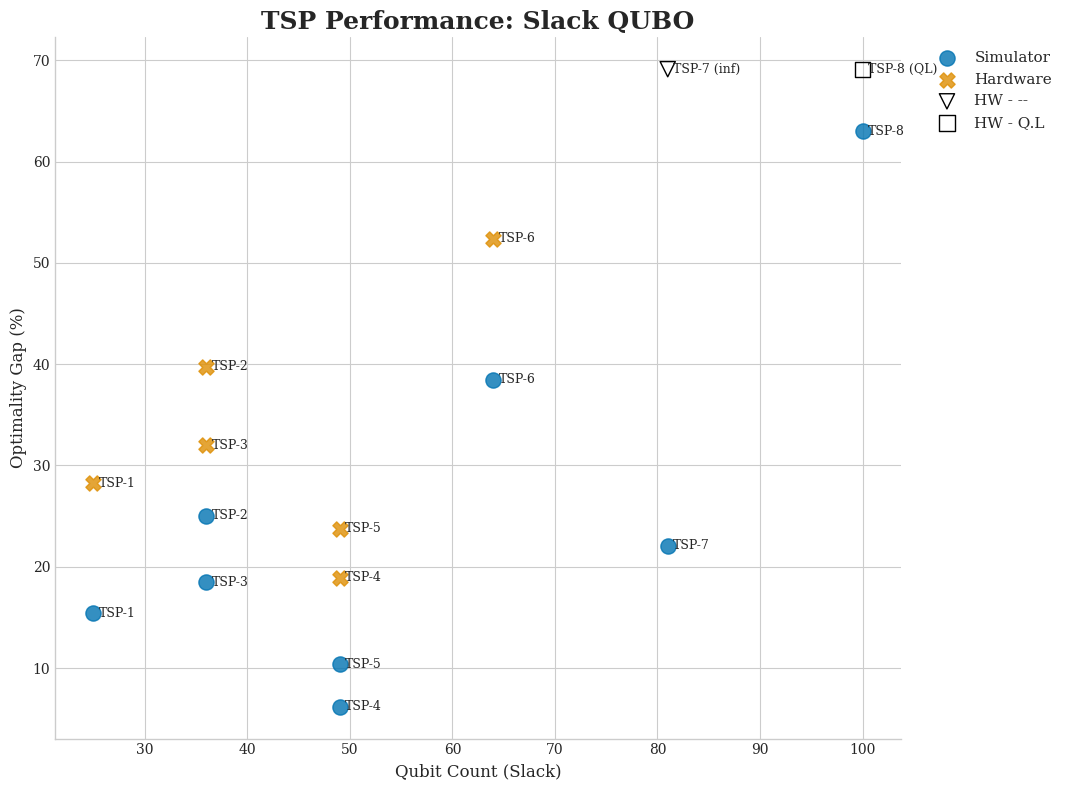}
        \caption{Slack QUBO Method}
        \label{fig:tsp-slack}
    \end{subfigure}
    \hfill
    \begin{subfigure}[b]{0.48\linewidth}
        \includegraphics[width=\linewidth]{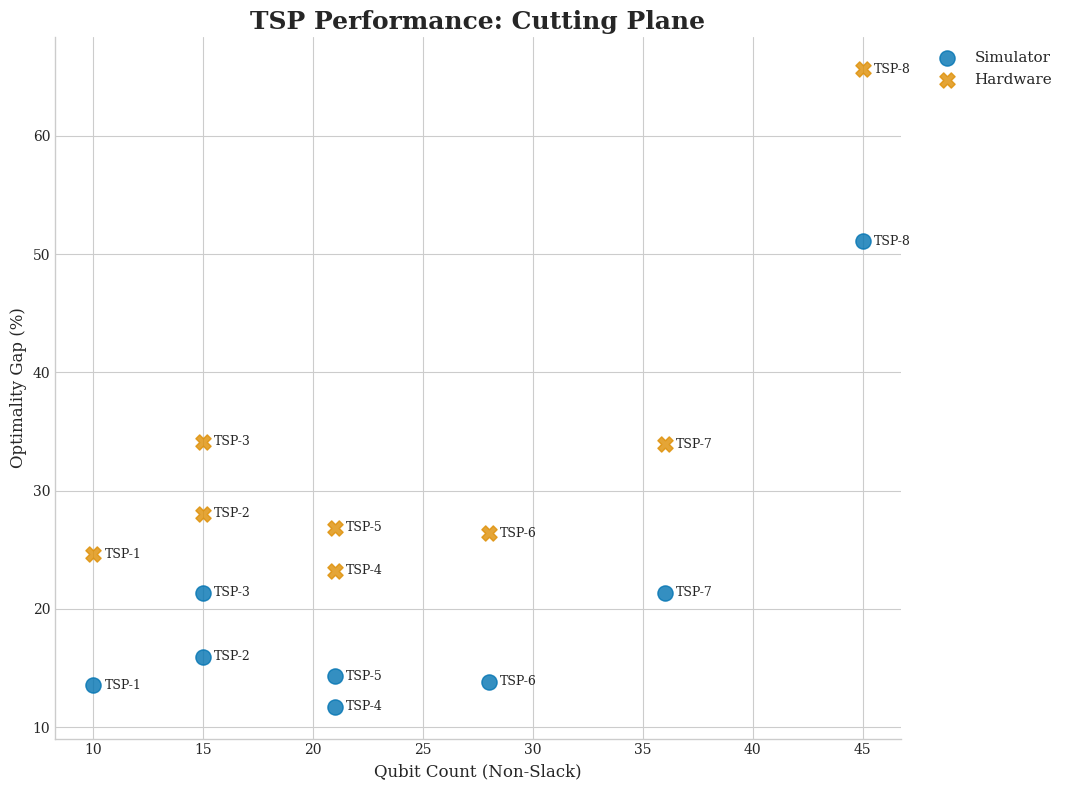}
        \caption{Cutting Plane Method}
        \label{fig:tsp-cp}
    \end{subfigure}
    
    \vspace{1em}
    
    \begin{subfigure}[b]{0.48\linewidth}
        \includegraphics[width=\linewidth]{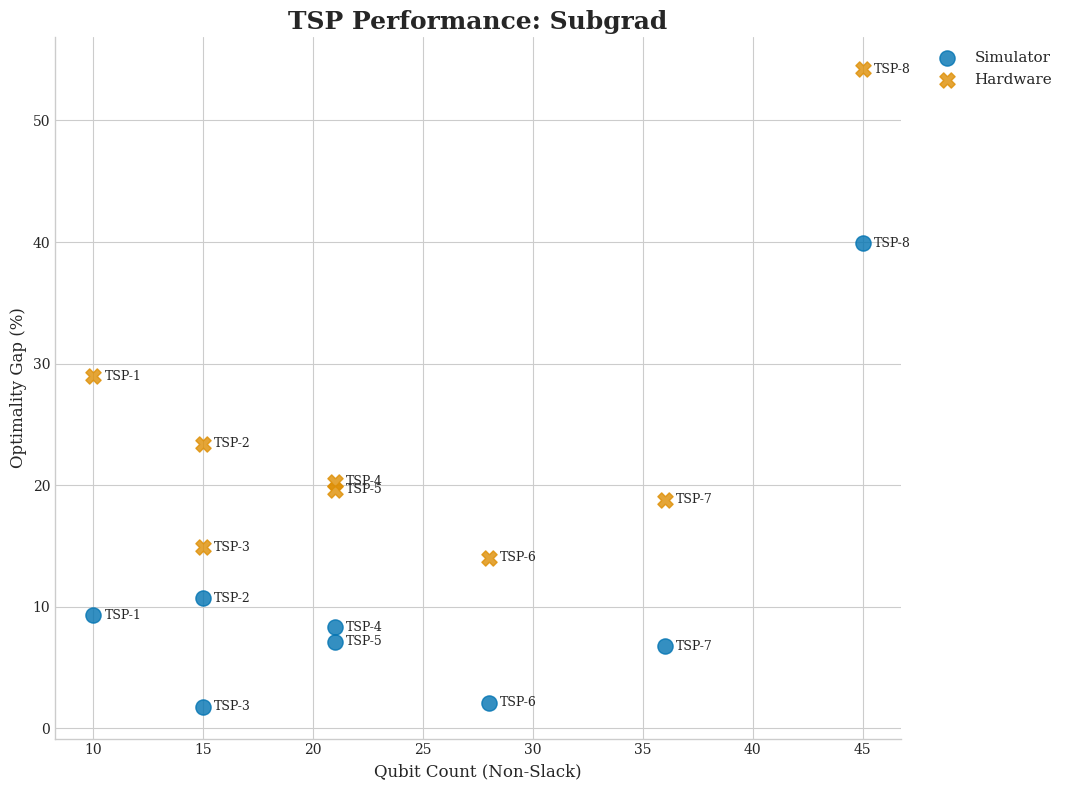}
        \caption{Subgradient Method}
        \label{fig:tsp-subgrad}
    \end{subfigure}
    \hfill
    \begin{subfigure}[b]{0.48\linewidth}
        \includegraphics[width=\linewidth]{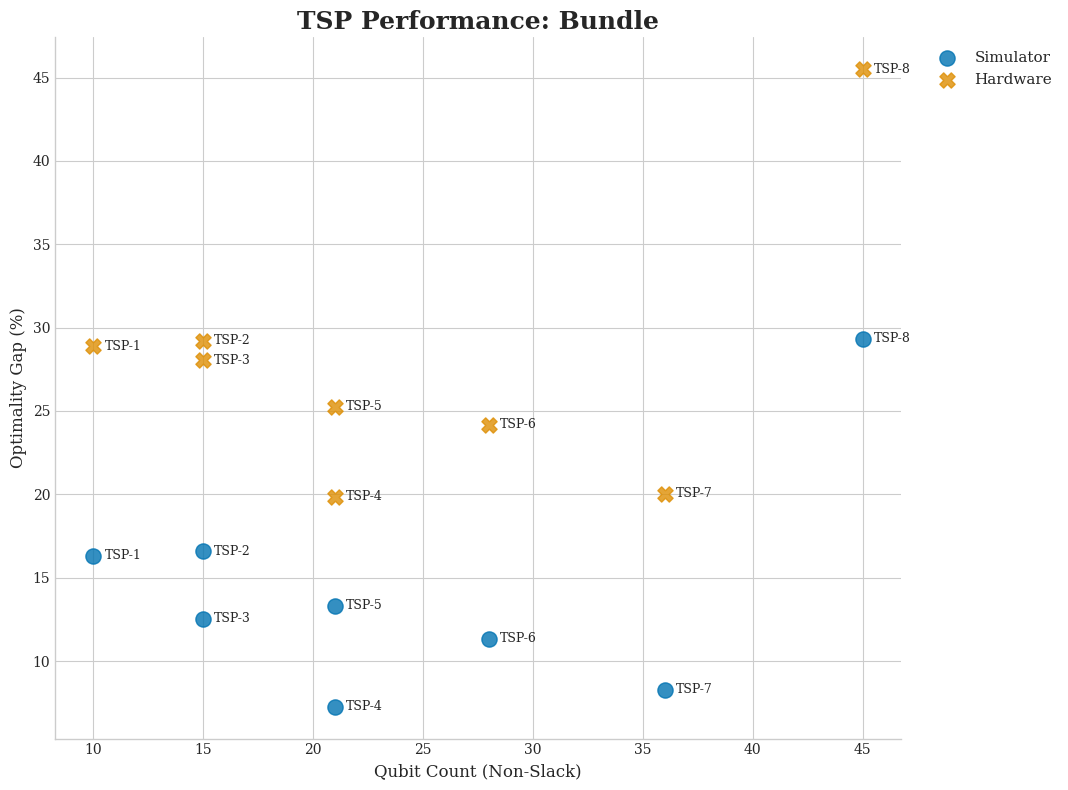}
        \caption{Bundle Method}
        \label{fig:tsp-bundle}
    \end{subfigure}
    
    \vspace{1em}
    
    \begin{subfigure}[b]{0.48\linewidth}
        \includegraphics[width=\linewidth]{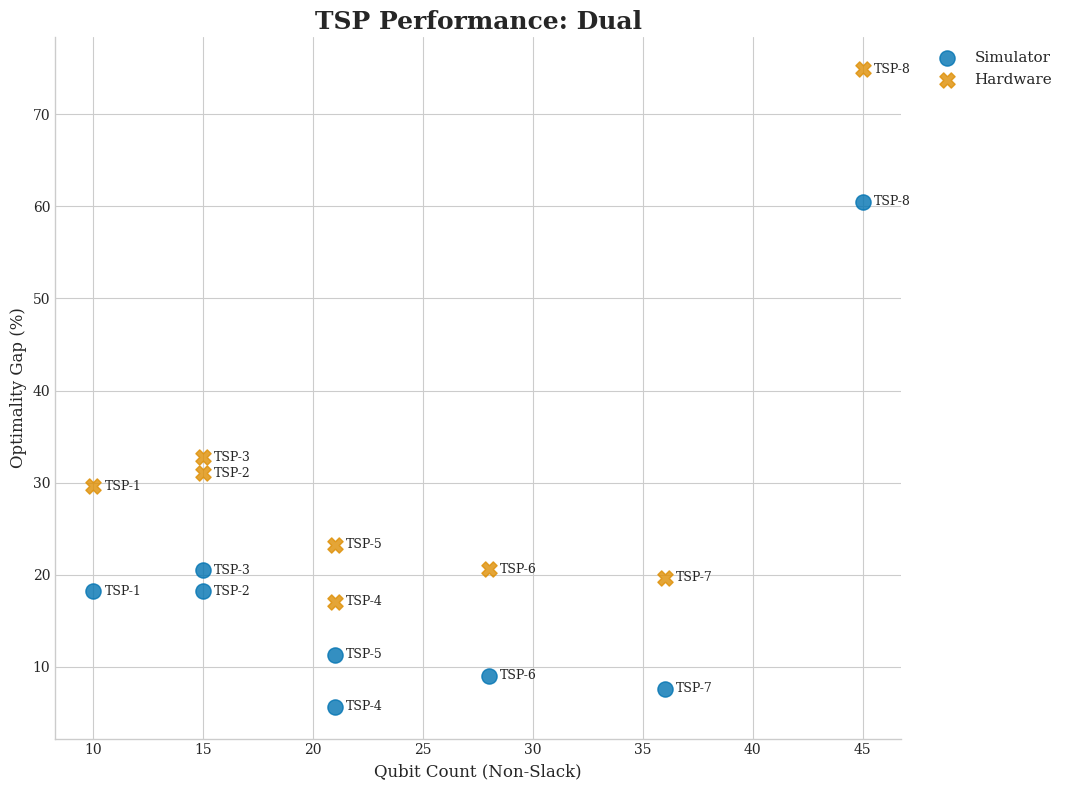}
        \caption{Dual Method}
        \label{fig:tsp-dual}
    \end{subfigure}
    
    \caption{Performance analysis of various optimization methods on the TSP instances. Each plot compares simulator and hardware results, showing the optimality gap versus the required number of qubits.}
    \label{fig:tsp-results}
\end{figure*}

\begin{figure*}[htbp!]
    \centering
    \begin{subfigure}[b]{0.48\linewidth}
        \includegraphics[width=\linewidth]{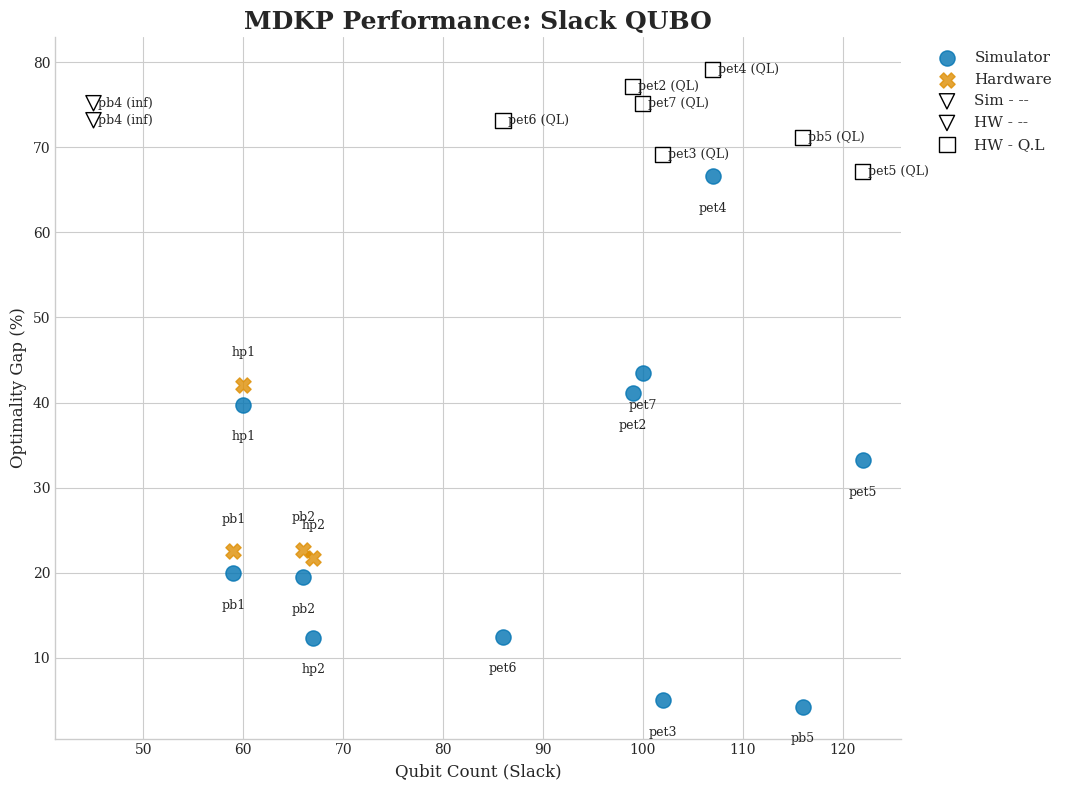}
        \caption{Slack QUBO Method}
        \label{fig:mdkp-slack}
    \end{subfigure}
    \hfill
    \begin{subfigure}[b]{0.48\linewidth}
        \includegraphics[width=\linewidth]{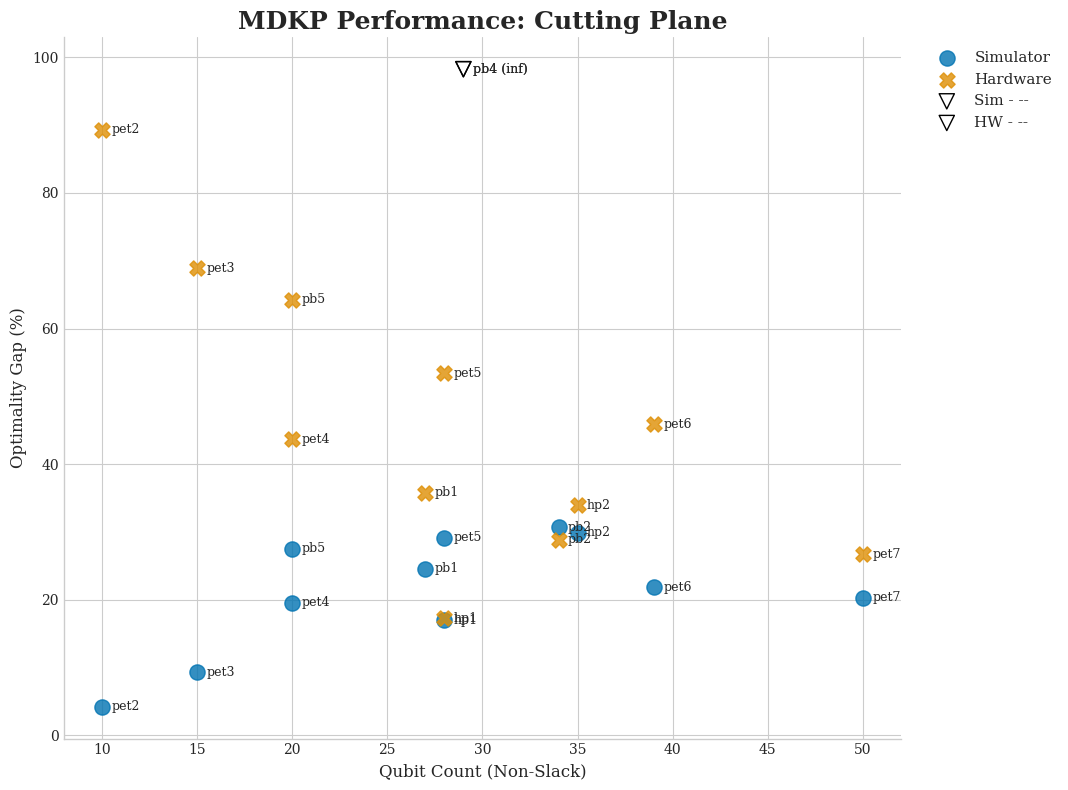}
        \caption{Cutting Plane Method}
        \label{fig:mdkp-cp}
    \end{subfigure}

    \vspace{1em}

    \begin{subfigure}[b]{0.48\linewidth}
        \includegraphics[width=\linewidth]{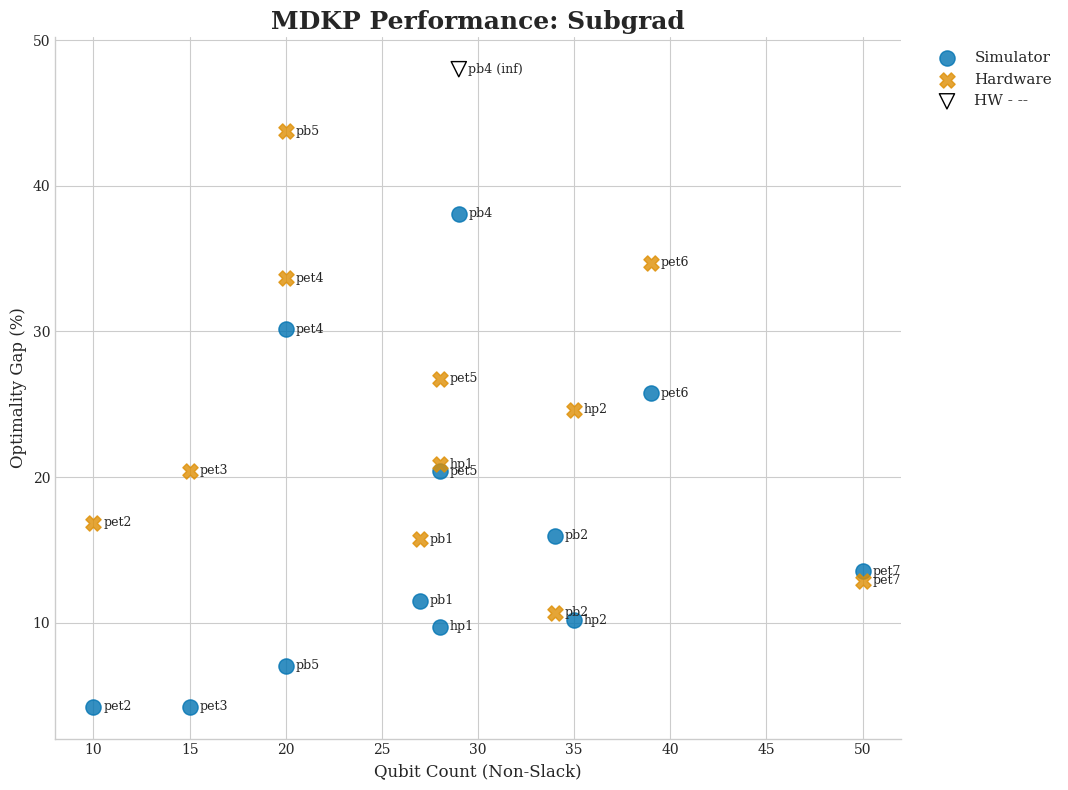}
        \caption{Subgradient Method}
        \label{fig:mdkp-subgrad}
    \end{subfigure}
    \hfill
    \begin{subfigure}[b]{0.48\linewidth}
        \includegraphics[width=\linewidth]{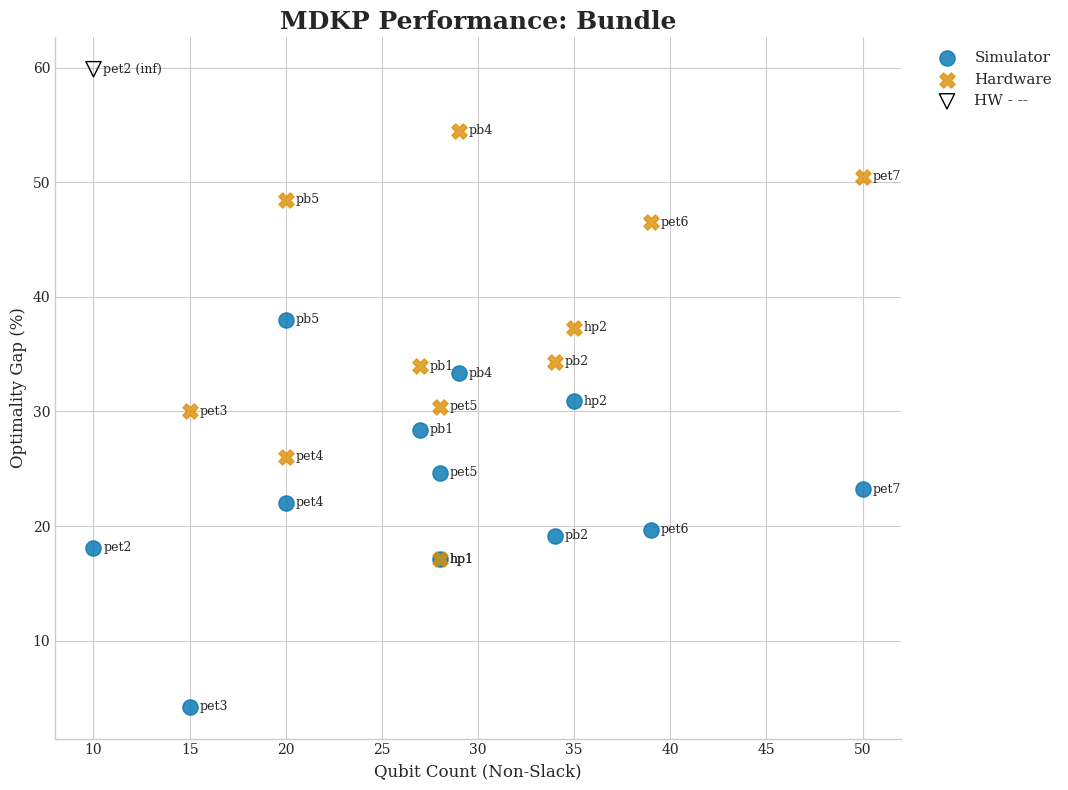}
        \caption{Bundle Method}
        \label{fig:mdkp-bundle}
    \end{subfigure}

    \vspace{1em}

    \begin{subfigure}[b]{0.48\linewidth}
        \includegraphics[width=\linewidth]{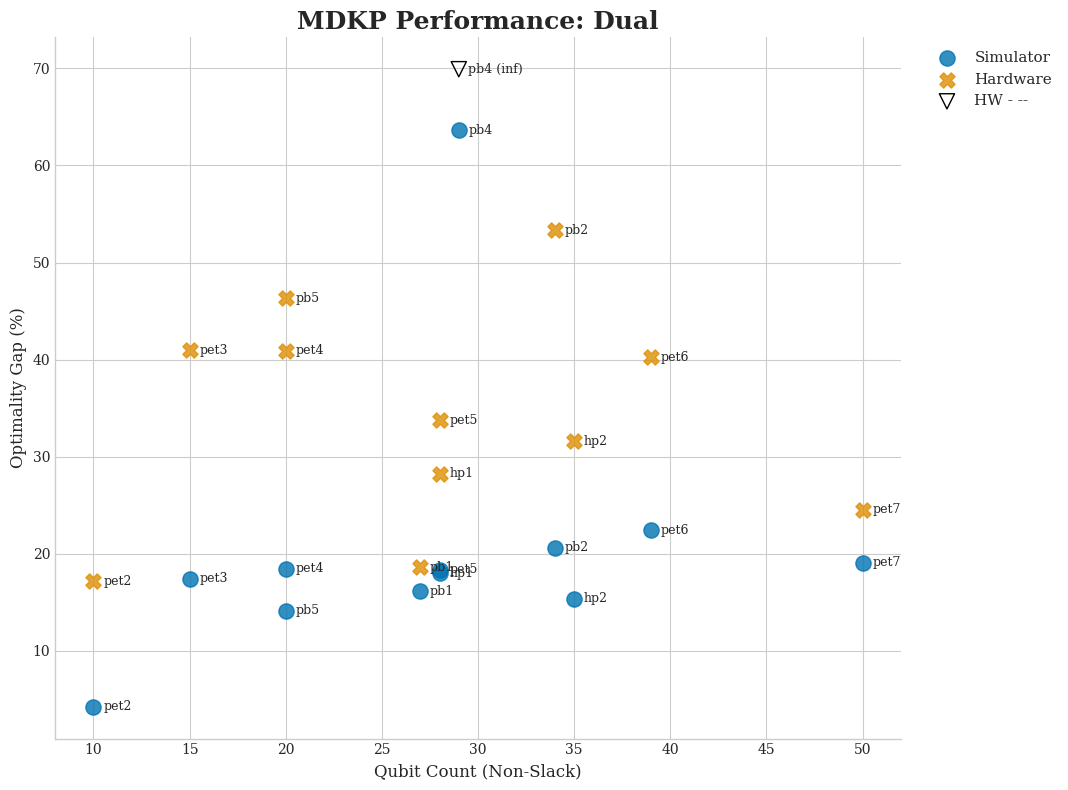}
        \caption{Dual Method}
        \label{fig:mdkp-dual}
    \end{subfigure}
    \hfill
    \begin{subfigure}[b]{0.48\linewidth}
        \includegraphics[width=\linewidth]{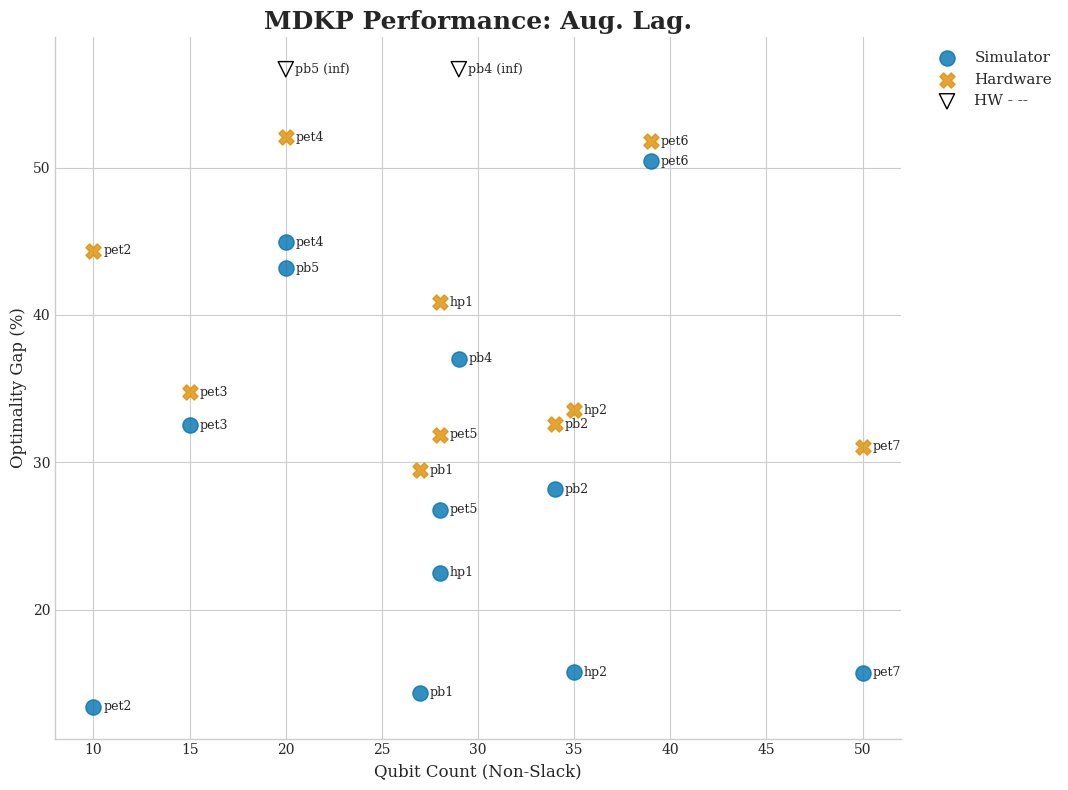}
        \caption{Augmented Lagrangian}
        \label{fig:mdkp-aug-lag}
    \end{subfigure}

    \caption{Performance analysis of various optimization methods on the MDKP instances. Each plot compares simulator and hardware results, showing the optimality gap versus the required number of qubits.}
    \label{fig:mdkp-results}
\end{figure*}

\begin{figure*}[htbp!]
    \centering
    \begin{subfigure}[b]{0.48\linewidth}
        \includegraphics[width=\linewidth]{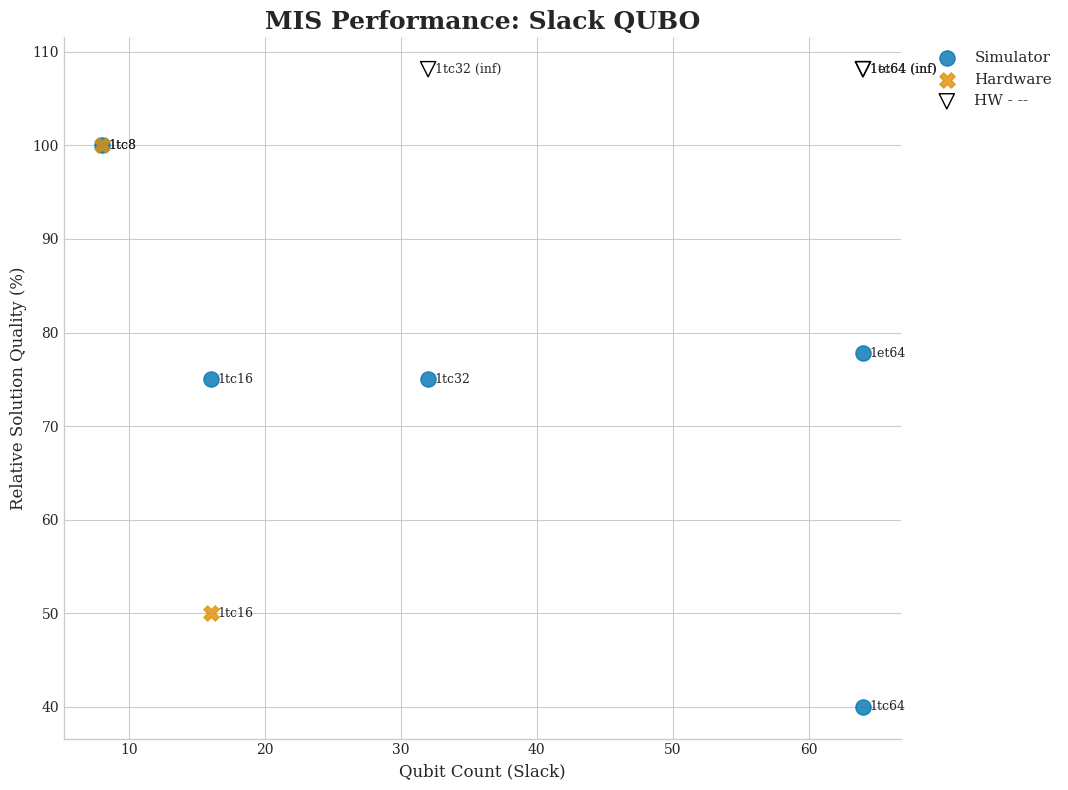}
        \caption{Slack QUBO Method}
        \label{fig:mis-slack}
    \end{subfigure}
    \hfill
    \begin{subfigure}[b]{0.48\linewidth}
        \includegraphics[width=\linewidth]{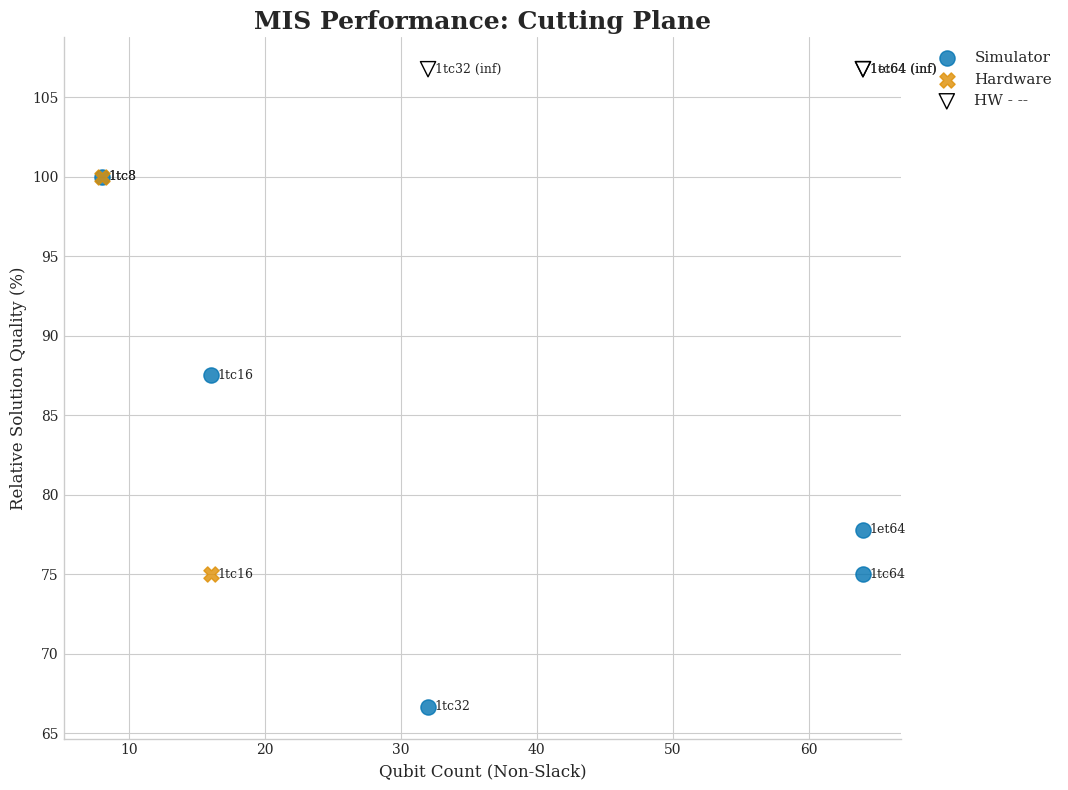}
        \caption{Cutting Plane Method}
        \label{fig:mis-cp}
    \end{subfigure}

    \vspace{1em}

    \begin{subfigure}[b]{0.48\linewidth}
        \includegraphics[width=\linewidth]{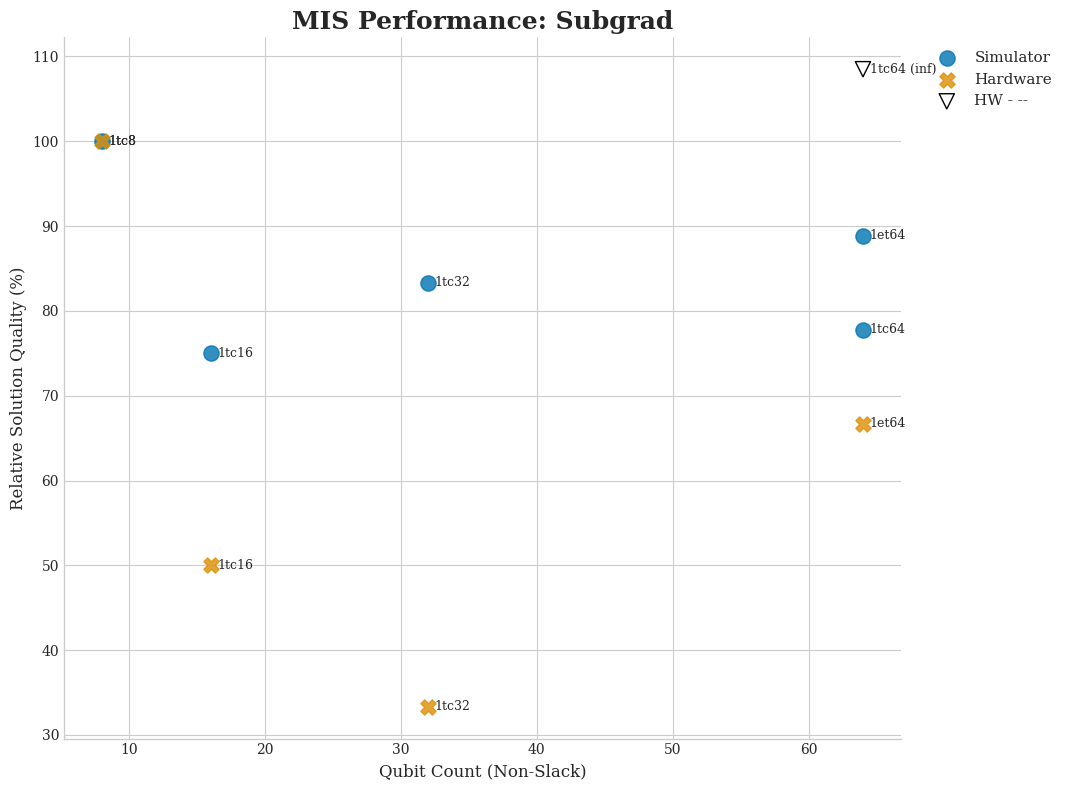}
        \caption{Subgradient Method}
        \label{fig:mis-subgrad}
    \end{subfigure}
    \hfill
    \begin{subfigure}[b]{0.48\linewidth}
        \includegraphics[width=\linewidth]{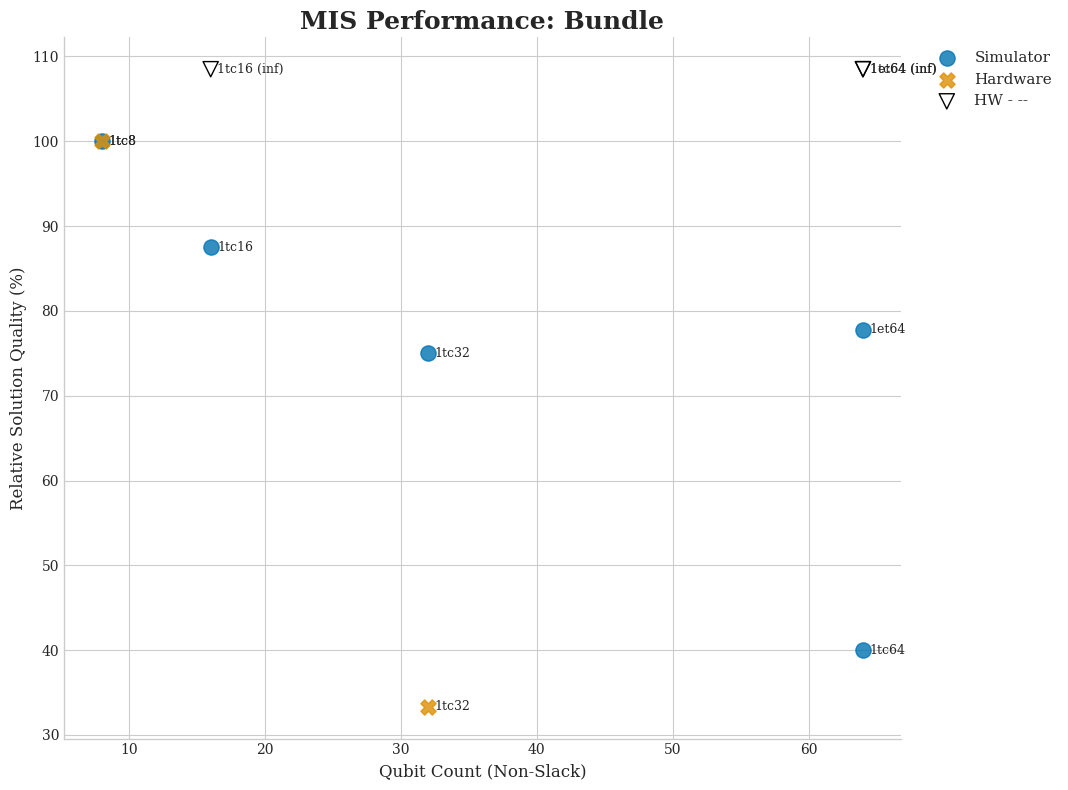}
        \caption{Bundle Method}
        \label{fig:mis-bundle}
    \end{subfigure}

    \vspace{1em}

    \begin{subfigure}[b]{0.48\linewidth}
        \includegraphics[width=\linewidth]{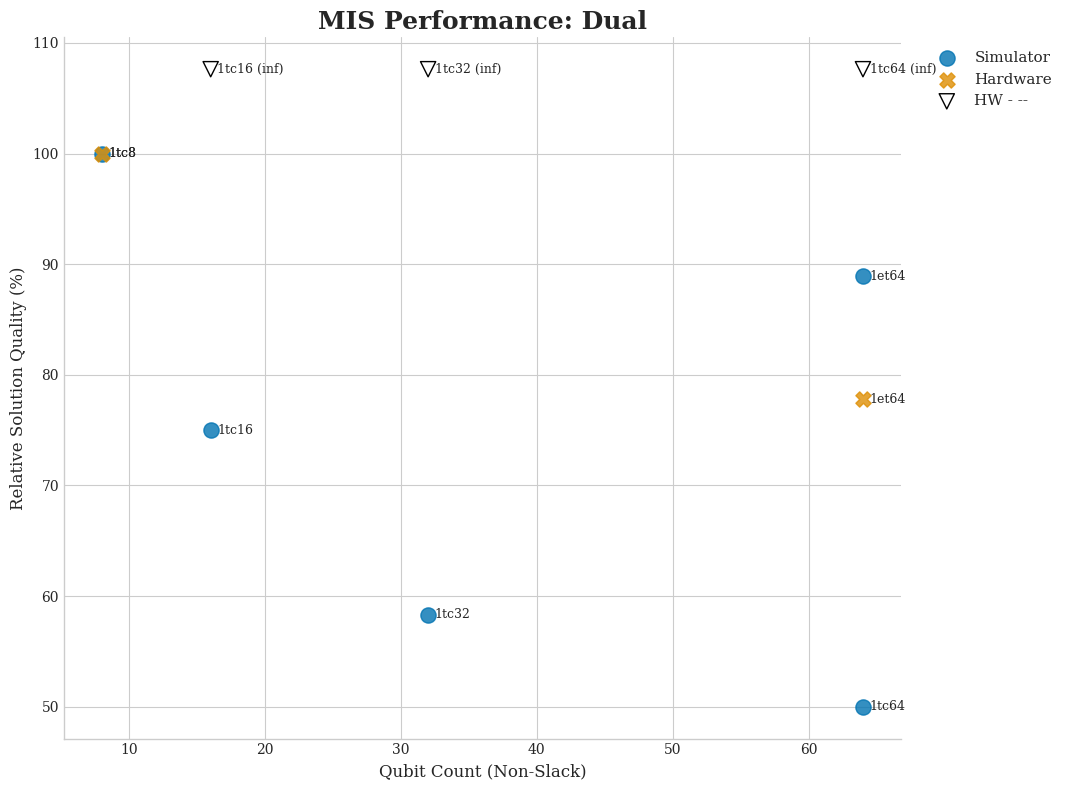}
        \caption{Dual Method}
        \label{fig:mis-dual}
    \end{subfigure}
    \hfill
    \begin{subfigure}[b]{0.48\linewidth}
        \includegraphics[width=\linewidth]{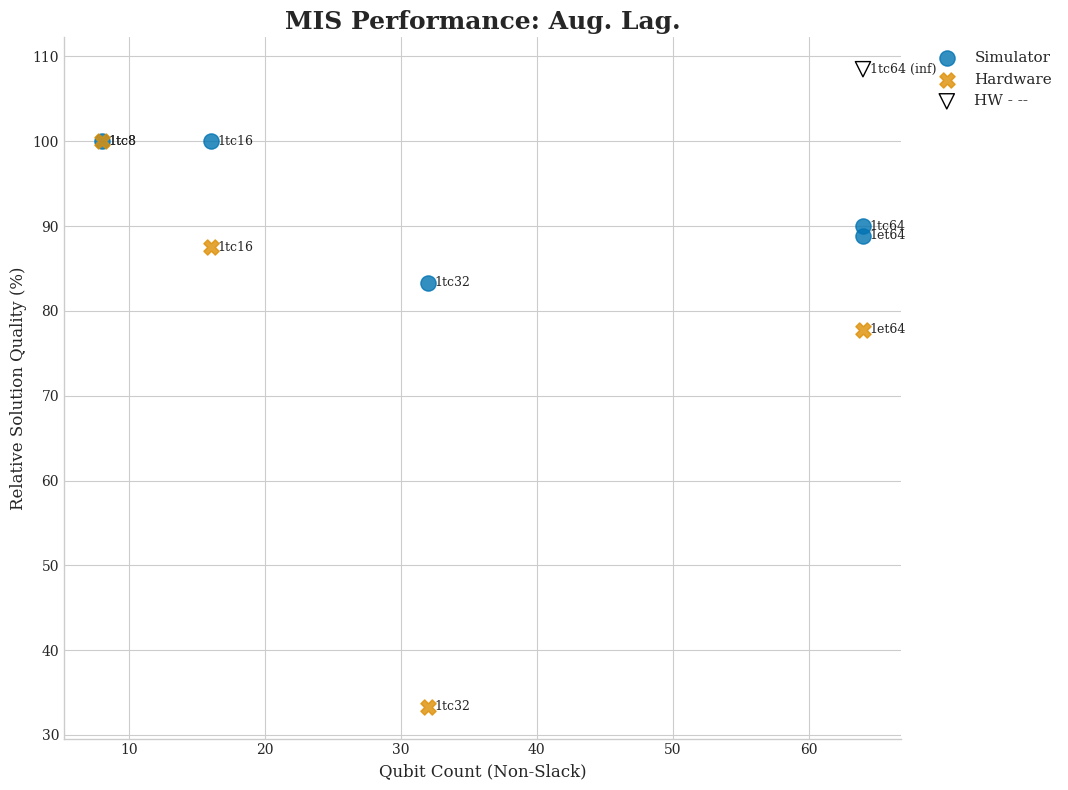}
        \caption{Augmented Lagrangian}
        \label{fig:mis-aug-lag}
    \end{subfigure}

    \caption{Performance analysis of various optimization methods on the MIS instances. Each plot compares simulator and hardware results, showing the relative solution quality versus the required number of qubits.}
    \label{fig:mis-results}
\end{figure*}

\subsubsection*{Visualization of Optimization Trends}

Figures~\ref{fig:tsp-results}, \ref{fig:mdkp-results}, and \ref{fig:mis-results} illustrate the optimization gap trends across TSP, MDKP, and MIS instances respectively, using various Lagrangian-based optimization techniques. These scatter plots map the relationship between the number of qubits used and the achieved solution quality for each method.

They also present side-by-side comparisons of simulator and hardware-executed results for TSP, MDKP, and MIS, respectively. In Figure~\ref{fig:tsp-results}, Subgradient and Dual methods consistently achieve lower optimality gaps across TSP instances, especially at lower qubit counts, underscoring their effectiveness in constraint handling. Figure~\ref{fig:mdkp-results} shows that for MDKP, slack-free approaches such as Subgradient and Bundle outperform slack-based QUBO formulations, offering better solution quality and reduced qubit requirements. For MIS, as shown in Figure~\ref{fig:mis-results}, the Augmented Lagrangian and Subgradient methods deliver strong performance, achieving Relative Solution Quality (RSQ) exceeding 90\% on multiple instances.

These visual patterns reinforce our earlier findings from Tables~\ref{tab:tsp-simulator-table}--\ref{tab:mis-simulator-table}.

\subsubsection*{Runtime Breakdown}

To assess the runtime efficiency of our approach, we distinguish between two components of the total time: the \textit{classical subroutine time}, which accounts for the time spent in parameter determination steps (such as dual averaging, subgradient updates, bundle methods, or cutting-plane strategies), and the \textit{quantum processing time}, which includes quantum circuit execution and post-processing (e.g., solution decoding or cost evaluation).

\begin{figure*}[ht]
    \centering
    \begin{subfigure}[t]{0.48\linewidth}
        \centering
        \includegraphics[width=\linewidth]{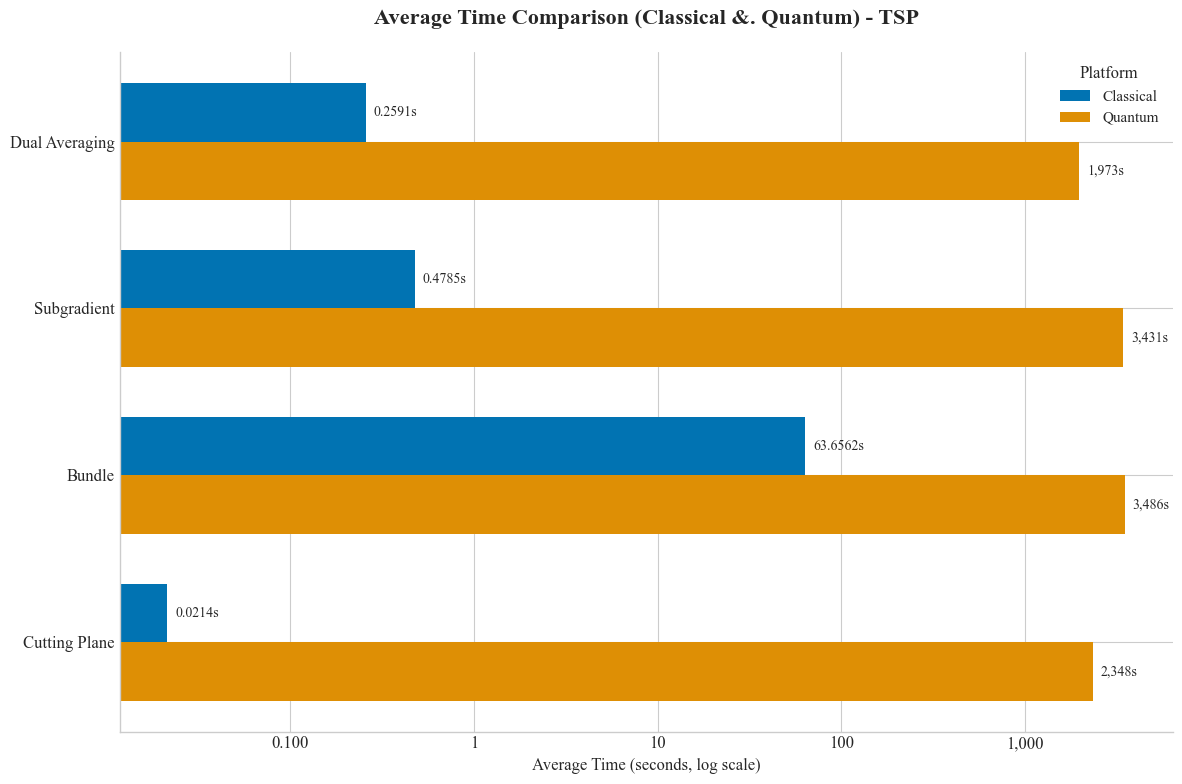}
        \caption{
        \textbf{TSP:} Classical (blue) and quantum (orange) average time per method. Times are on a log scale.
        }
        \label{fig:tsp-time}
    \end{subfigure}
    \hfill
    \begin{subfigure}[t]{0.48\linewidth}
        \centering
        \includegraphics[width=\linewidth]{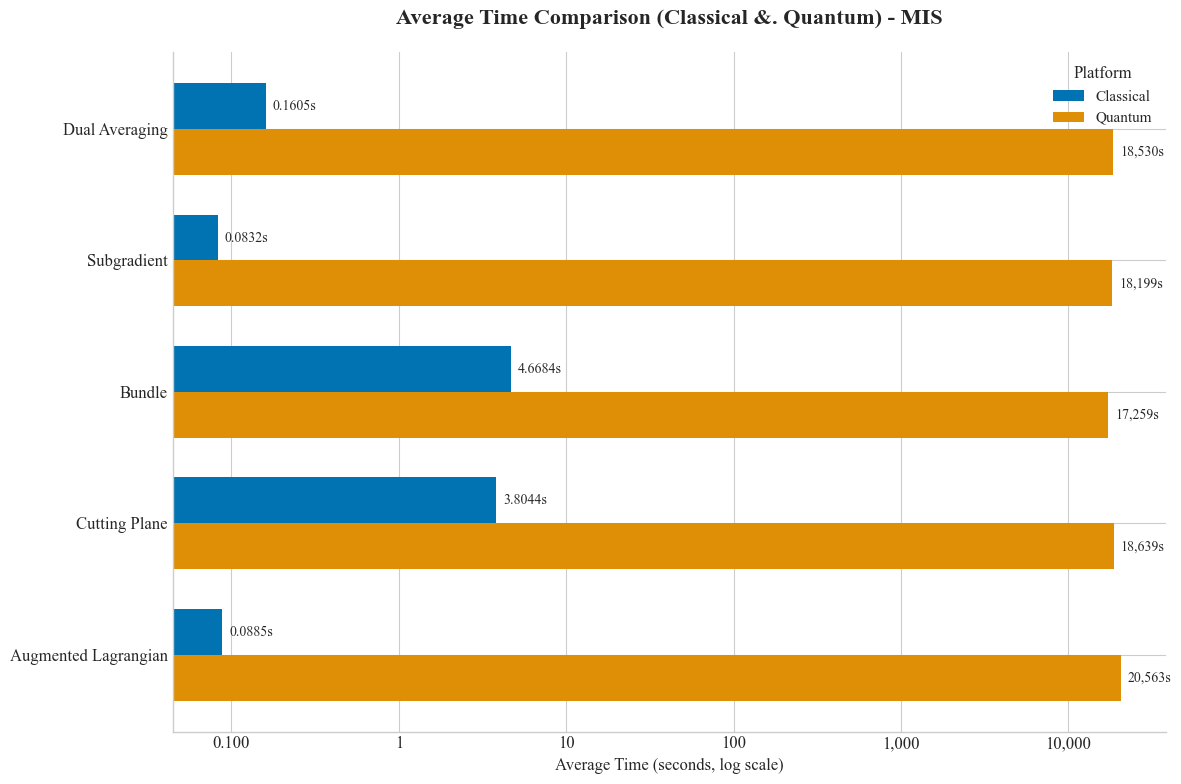}
        \caption{
        \textbf{MIS:} Average runtime comparison. Classical time reflects subroutines, quantum time includes circuit execution.
        }
        \label{fig:mis-time}
    \end{subfigure}
    \caption{
    \textbf{Average time comparison (Classical vs. Quantum) for TSP and MIS.}
    Runtime is separated into classical subroutine time and quantum execution time for various optimization methods. All values are shown in seconds on a logarithmic scale.
    }
    \label{fig:tsp-mis-time}
\end{figure*}

\begin{figure*}[ht]
    \centering
    \includegraphics[width=0.8\linewidth]{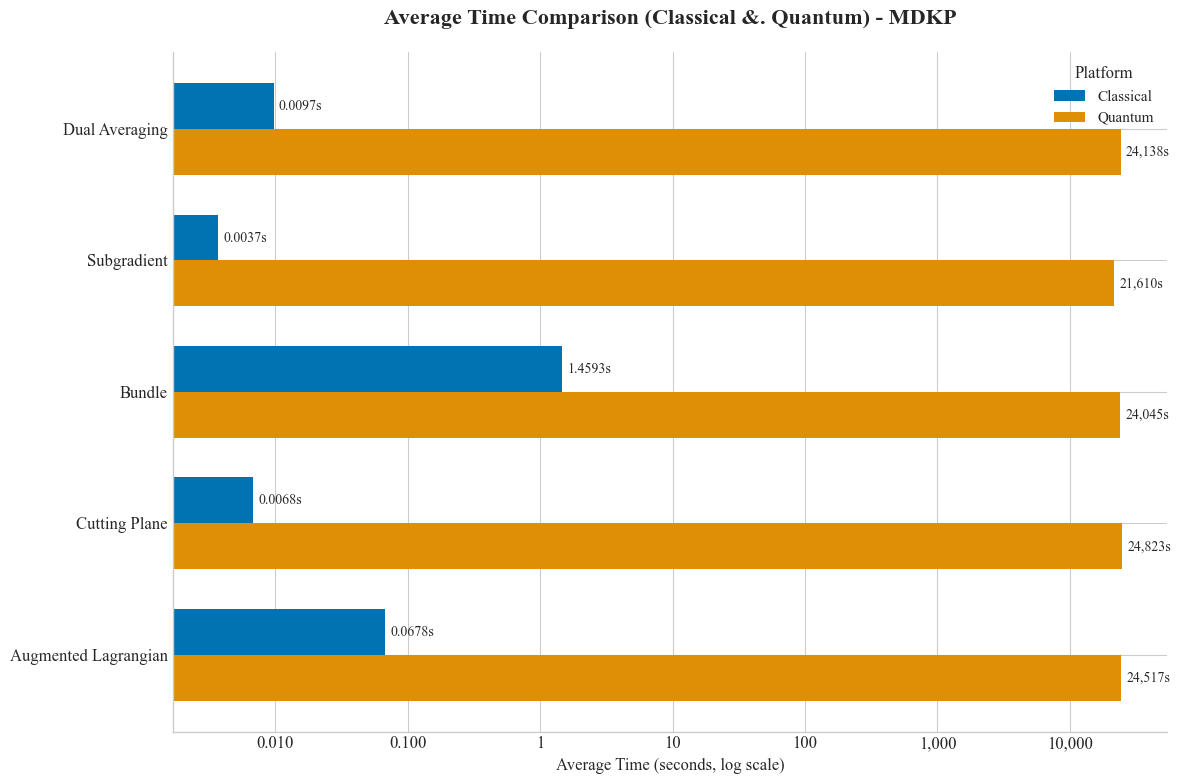}
    \caption{
    \textbf{Average time comparison (Classical vs. Quantum) for MDKP.}
    The bars highlight the disparity between efficient classical parameter subroutines and quantum execution times. All runtimes are in seconds (log scale).
    }
    \label{fig:mdkp-time}
\end{figure*}

Figure~\ref{fig:tsp-mis-time} and \ref{fig:mdkp-time} reports the average time required by each method across three benchmark problem classes: \textbf{TSP}, \textbf{MIS}, and \textbf{MDKP}. The time axis is shown on a logarithmic scale to accommodate the significant disparity between classical and quantum runtimes. The classical bars (in blue) highlight the low computational overhead of traditional parameter updates, while the quantum bars (in orange) emphasize the considerable execution cost of quantum subroutines. These plots provide a comparative view of the performance bottlenecks across methods and help identify which classical strategies offer the best trade-off between runtime and solution quality.

\section{\label{sec:discussion} Discussion and Conclusion}

\subsection{\textbf{Insights from Experimental Results}}

Our empirical study provides insights into the performance trade-offs of different constraint-handling strategies across three combinatorial problems: TSP, MDKP, and MIS. We focus on the dual goals of minimizing qubit requirements and maintaining competitive solution quality under quantum simulation and hardware settings.

\subsubsection{\textbf{Qubit Efficiency and Resource Constraints}}

As observed in Tables~\ref{tab:tsp-simulator-table}--\ref{tab:mis-simulator-table}, the slack-based (S) formulations consistently require more qubits than their slack-free (NS) counterparts. This increase is due to the explicit introduction of auxiliary variables that handles inequality constraints. While slack-based encodings offer structural clarity, they rapidly inflate resource demands, often surpassing current hardware limits. In contrast, slack-free methods rely on Lagrangian or penalty-based encodings, achieving significant qubit savings without compromising the fidelity of constraint penalty in most cases.

Notably, multiple MDKP instances in Table~\ref{tab:mdkp-hardware-table}, and one in TSP instances Table~\ref{tab:tsp-hardware-table} are marked as Q.L (qubit limit), particularly for slack-based encodings. This reinforces the practical limitations of current quantum hardware, where qubit count and connectivity remain primary bottlenecks. For example, instance \texttt{pb5} required 116 qubits under the slack-based formulation, making it infeasible to execute on Rigetti’s Ankaa‑3 QPU.

\subsubsection{\textbf{Optimization Method Comparison}}

Across all three problem classes, MDKP, MIS, and TSP-we observe consistent trends regarding the effectiveness of various optimization strategies. Direct QUBO-based formulations, especially the slack-based QUBO, often yield the highest optimization gaps. This is primarily due to the rigid nature of static penalty terms, which can either inadequately enforce constraints or excessively penalize feasible solutions. For instance, in MDKP instance \texttt{pb2}, the slack QUBO yields a 22.70\% gap on hardware, whereas the Subgradient method reduces it to just 10.70\%. Similarly, for TSP instance \texttt{TSP-3}, the gap decreases from 18.45\% (slack QUBO) to 1.76\% using Subgradient optimization.

Iterative methods such as Subgradient and Bundle consistently outperform the direct QUBO baseline. These methods leverage dynamic updates of dual variables or gradient history, enabling more stable convergence to feasible and near-optimal solutions. In the MIS setting, where optimal solutions are known, these methods yield high Relative Solution Quality (RSQ) scores, frequently exceeding 90\%. For example, in instance \texttt{1tc64}, the Augmented Lagrangian method reaches 90\% RSQ, compared to just 40\% under the slack QUBO formulation.

The Cutting Plane method also shows competitive performance in MDKP and TSP, but its iterative addition of constraints increases complexity. The Augmented Lagrangian approach performs well in MIS and MDKP, but its performance is highly sensitive to penalty coefficient tuning. For instance, in \texttt{pet6}, tuning failure led to an optimization gap of over 50\%.

For TSP, we did not implement the Augmented Lagrangian formulation. Instead, we applied a Held-Karp style Lagrangian relaxation tailored to the tour-structure constraints of TSP. This choice offered a more natural way to handle sub-tour elimination and edge consistency, while maintaining a compact qubit encoding.

\subsubsection{\textbf{Unified Observations Across Problem Types}}

When comparing performance across MDKP, MIS, and TSP, slack-free methods (NS) consistently demonstrate superior qubit efficiency and competitive solution quality. The Subgradient method stands out as a robust approach across all problems, achieving low optimization gaps in TSP, strong convergence in MDKP, and high RSQ in MIS. The Bundle method follows closely, particularly for problems with complex constraint interactions.

In contrast, slack-based QUBO formulation, despite being conceptually simple-scale poorly in both qubit requirements and optimization quality. The dependence on static penalty terms makes them difficult to tune and often impractical for hardware execution. Augmented Lagrangian methods, where applicable, offer a strong balance between feasibility and solution quality, but require careful parameter tuning.

Taken together, these results suggest that hybrid methods, combining classical constraint handling with quantum optimization, are essential for solving real-world combinatorial problems on near-term quantum devices. Efficient encodings, dynamic dual updates, and problem-specific relaxations (such as Held-Karp for TSP) emerge as key design choices in constructing scalable quantum optimization workflows.

\subsection{\textbf{Practical Implications for Quantum Hardware}}

\subsubsection{\textbf{Scalability on Near-Term Devices}}

The qubit limit annotations in Table~\ref{tab:mdkp-hardware-table} illustrate the scalability challenges of NISQ-era devices. Even moderate-size MDKP instances push hardware constraints, particularly under slack-based formulations. The NS approach, by minimizing qubit footprint, offers a viable path forward. Its compatibility with shallow circuits makes it amenable to current quantum processors with limited coherence times and noisy gate operations.

\subsubsection{\textbf{Trainability and Hybrid Optimization}}

Quantum solvers like VQE rely on parameterized circuits, whose trainability is adversely affected by increased qubit count and depth. Slack-free encodings help mitigate these issues, enabling shallower circuits and improved optimization dynamics. Integrating classical optimization routines (e.g., subgradient updates, dual averaging) in a hybrid loop with quantum circuits significantly improves convergence, especially under noisy execution.

\subsubsection{\textbf{Industry-Relevant Applications}}

Our findings suggest that domains such as logistics, resource allocation, and portfolio optimization, which can be modeled via knapsack or independent set structures, can benefit from quantum optimization in the near term. However, realizing this potential requires careful attention to encoding strategies, qubit efficiency, and hybrid scheme design. Slack-free formulations stand out as particularly promising due to their hardware feasibility and flexibility in enforcing constraints.

\subsection{\textbf{Summary of Key Findings}}

\begin{itemize}
    \item Slack-free formulations consistently reduce qubit requirements, often achieving comparable or better solution quality than slack-based approaches.
    \item Direct QUBO formulations underperform in many cases unless penalties are meticulously tuned.
    \item Adaptive methods such as Subgradient and Bundle show robust performance across problem types and execution settings.
    \item Augmented Lagrangian methods are effective but require careful hyperparameter tuning to avoid instability.
    \item Qubit limitations remain a significant barrier for large instances, making compact formulations essential on NISQ devices.
    \item Slack-free encodings are especially suited for real-world deployment, offering a path to practical quantum advantage via hybrid algorithms.
\end{itemize}

While classical solvers remain superior for small- to medium-scale instances, our experiments emphasize the complementary role of classical techniques in quantum optimization. As quantum hardware continues to evolve, efficient formulations and hybrid designs will be key to unlocking their full potential in combinatorial optimization.

\subsection{ \textbf{Future Directions}}

Despite the promising results, several challenges remain. Future work should explore:

\begin{itemize}
    \item  \textbf{Hybrid Quantum-Classical Methods}: Enhancing feasibility via  dynamic penalty updates and  constraint violation monitoring.
    \item  \textbf{Benchmarking on Alternative Quantum Architectures}: Evaluating  neutral-atom quantum processors (e.g.,  QuEra’s Aquila) to assess formulation performance across different hardware paradigms.
    \item  \textbf{Scaling to Larger Problem Instances}: Investigating the behavior of NS formulations on larger instances and related combinatorial problems.
    \item  \textbf{Improving Quantum Constraint Handling}: Developing  quantum-enhanced feasibility correction methods to further improve slack-free formulations.
\end{itemize}

\subsection{\textbf{ Conclusion}}

This study highlights a fundamental trade-off in quantum optimization:  while slack-based formulations can separate feasible and infeasible solutions, they require additional qubits, making them impractical for current NISQ devices. In contrast,  slack-free formulations offer competitive performance with reduced qubit requirements, making them a more viable choice for near-term quantum optimization. The results contribute to the growing effort in designing scalable quantum optimization frameworks for real-world combinatorial problems.

As quantum hardware continues to evolve, techniques that minimize qubit usage, such as the slack-free formulations will become increasingly critical for addressing larger-scale combinatorial optimization challenges. The results presented here, obtained on today’s limited devices, establish a vital foundation for future efforts to tackle industry-scale problems as more powerful and error-resilient quantum computers emerge. By reducing resource overhead and incorporating classical insights into constraint handling, our approach not only enhances current performance but also paves the way toward realizing quantum advantage in practical, large-scale applications.

\section{Acknowledgement}

This research is supported by the National Research Foundation, Singapore under its Quantum Engineering Programme 2.0 (NRF2021-QEP2-02-P01).

\bibliography{references}

\newpage

\appendix

\section{\label{app:algorithms}Algorithms}

\begin{figure}[h!]
    \begin{algorithmic}[1]
        \caption{Integration of Lagrangian Relaxation with Quantum Optimization}
        \label{alg:lagrangian_integration}
        \Require MKP instance data, initial Lagrange multipliers $\lambda^{(0)}$, maximum iterations $K$, tolerance $\epsilon$.
        \Ensure Best Lagrange multipliers $\lambda^*$ and final QUBO solution $x^*$.
        \State \textbf{Initialize} $\lambda \gets \lambda^{(0)}$.
        \For{$k = 1, 2, \dots, K$}
            \State \textbf{Solve Lagrangian Subproblem:} For each item $i$, set
            $$
            x_i^*(\lambda) =
            \begin{cases}
            1, & \text{if } p_i - \sum_{j=1}^{m} \lambda_j w_{ji} > 0,\\[1ex]
            0, & \text{otherwise}.
            \end{cases}
            $$
            \State \textbf{Compute} the dual function value:
            $$
            g(\lambda) = \sum_{j=1}^{m} \lambda_j b_j + \sum_{i=1}^{n} \max\{0,\, p_i - \sum_{j=1}^{m} \lambda_j w_{ji}\}.
            $$
            \State \textbf{Calculate} the subgradient for each constraint $j$:
            $$
            s_j = b_j - \sum_{i=1}^{n} w_{ji}\, x_i^*(\lambda).
            $$
            \If{$\left\| s \right\| < \epsilon$}
                \State \textbf{break}
            \EndIf
            \State \textbf{Update} the Lagrange multipliers:
            $$
            \lambda_j \gets \max\{0,\, \lambda_j - \alpha_k\, s_j\}, \quad \forall j,
            $$
            where $\alpha_k$ is the step size at iteration $k$.
        \EndFor
        \State \textbf{Set} $\lambda^* \gets \lambda$.
        \State \textbf{Formulate QUBO:} Insert $\lambda^*$ into the Lagrangian-relaxed objective to define
        $$
        \min_{x \in \{0,1\}^n} -\Biggl( \sum_{i=1}^{n}\Bigl[p_i - \sum_{j=1}^{m} \lambda^*_j w_{ji}\Bigr] x_i + \sum_{j=1}^{m} \lambda^*_j b_j \Biggr).
        $$
        \State \textbf{Convert} the formulation to a QUBO and solve it using a quantum solver (e.g., via VQE).
        \State \textbf{Return} the final QUBO solution $x^*$ along with $\lambda^*$.
    \end{algorithmic}
\end{figure}

\end{document}